\documentclass[5p]{elsarticle}

\usepackage[english]{babel}
\usepackage[latin1]{inputenc}
\usepackage[T1]{fontenc}      
\usepackage{units}
\usepackage{setspace}
\usepackage{color}
\usepackage{amsmath}
\usepackage{psfrag}

\biboptions{sort&compress}

\newcommand{\TEM}{TEM$_{00} $ } %

\newcommand{\muu}{m_{\gamma^{\prime}}}

\usepackage{amssymb}

\journal{Nuc. Instrum. Meth.}

\begin{document}

\begin{frontmatter}

\title{
\hfill {\small DESY 09-058}\\
Resonant laser power build-up in ALPS \\ -- a ``light-shining-through-walls'' experiment --}

\author[a]{Klaus Ehret}
\author[b]{Maik Frede}
\author[a]{Samvel Ghazaryan}
\author[b]{Matthias Hildebrandt}
\author[a]{Ernst-Axel Knabbe}
\author[b]{Dietmar Kracht}
\author[a]{Axel~Lindner}
\author[a]{Jenny List}
\author[c]{Tobias Meier}
\author[a]{Niels Meyer}
\author[a]{Dieter Notz}
\author[a]{Javier Redondo}
\author[a]{Andreas Ringwald}
\author[d]{G\"unter~Wiedemann}
\author[c]{Benno Willke}

\address[a]{Deutsches Elektronen-Synchrotron DESY, Notkestra\ss e 85, D-22607 Hamburg, Germany}
\address[b]{Laser Zentrum Hannover e.V., Hollerithallee 8, D-30419 Hannover, Germany}
\address[c]{Max-Planck-Institute for Gravitational Physics, Albert-Einstein-Institute, and Institut f\"ur Gravitationsphysik, 
Leibniz Universit\"at, Hannover, Callinstraße 38, D-30167 Hannover, Germany}
\address[d]{Hamburger Sternwarte, Gojenbergsweg 112, D-21029 Hamburg, Germany}

\begin{abstract}
The ALPS collaboration runs a light-shining-through-walls (LSW) experiment to search for photon oscillations into ``weakly interacting sub-eV particles'' (WISPs) inside of a superconducting HERA dipole magnet at the site of DESY. 
In this paper we report on the first successful integration of a large-scale optical cavity to boost the available power for WISP production in this type of experiments.
The key elements are a frequency tunable narrow line-width continuous wave laser acting as the primary light source and an electronic feed-back control loop to stabilize the power build-up.
We describe and characterize our apparatus and demonstrate the data analysis procedures on the basis of a brief exemplary run. 

\end{abstract}

\end{frontmatter}

\newpage

\section{Introduction} \label{intro}

The standard model of elementary particles (SM) offers an accurate description of 
almost every phenomenon observed so far in particle physics. 
Despite this success, there are both theoretical and observational motivations to 
believe what the SM describes is just a small component of nature's complexity.
From the theoretical side, the SM suffers from naturalness, hierarchy and
arbitrariness problems and, moreover, it does not describe gravity. From the observational
viewpoint, cosmology and astrophysics claim that around 22\% of the universe's energy 
density is made of a yet unidentified type of non-baryonic ``dark matter'' and $74$\% of a 
yet more mysterious ``dark energy''.

Extensions of the SM usually introduce many new particles and symmetries that provide 
elegant solutions to our theoretical concerns and candidates for the ``dark sector''.
Interestingly enough, these extensions often include much more particles than known today.
On general grounds these particles could live in principle at any energy scale since 
new symmetries could protect their masses.

In particular nothing prohibits that light particles beyond the SM exist as long 
as they have no SM  charges, i.e. as long as they populate a ``hidden sector''.
Actually these ``hidden sectors'' arise quite naturally in string theory, our current best candidate
for a theory of quantum gravity, and they are required to break the hypothetical supersymmetry 
which would solve the hierarchy problem and might provide good dark matter candidates, the so-called 
WIMPs (for weakly interacting massive particles).

Low mass particles living in ``hidden sectors'' might still have very weak interactions with the SM
fields through radiative corrections involving massive ``mediator'' particles or through gravity. 
We name these particles WISPs (weakly%
\footnote{Here ``weak'' has to be understood in a broad sense, while in WIMP, it stands for electro-weak.} 
interacting sub-eV particles).
The feebleness of their interactions implies extremely few events in a typical experiment, 
making the luminosity and detector efficiency two crucial requirements for WISP searches.
On the other hand, having low masses, WISPs can potentially exhibit coherent interactions 
along \emph{macroscopic} distances boosting their production probabilities. 

Currently, the beams of SM particles with the highest luminosities and best coherence 
properties are \emph{laser} beams. 
Commercial lasers in the visible spectrum can easily reach output powers of several tens of Watts 
($\gtrsim 10^{19}$ photons per second) and can have coherence lengths of several kilometers.
Sensitive detectors with quantum efficiencies close $\sim 100\%$ are also available. 
Therefore, on general grounds, experimental searches for WISPs coupled to photons 
are the most favored.

Let us notice that these WISPs can have a sizable impact on cosmology  
and astrophysics.
Often not realized, they can be as good \emph{cold} dark matter  
candidates as WIMPs
despite having sub-eV masses, as it is the case for the axion~ 
\cite{Abbott:1982af}.
Stellar evolution is often accelerated by the emission of WISPs~ 
\cite{Raffelt:1996wa,Raffelt:1999tx,Redondo:2008aa,Gninenko:2008pz},
as it is by neutrinos, and photon-WISP interactions can modify the  
spectra of cosmic
radiation~\cite{Georgi:1983sy,DeAngelis:2002hm,Mirizzi:2005ng,Mirizzi:2007hr,Mirizzi:2009iz,Jaeckel:2008fi}. Indeed, WISPs could help in  
understanding a number of recent observations~\cite{Jaeckel:2008fi,DeAngelis:2007dy,Huh:2007zw,Isern:2008fs,Isern:2008nt,Burrage:2009mj,
Fairbairn:2009zi,Ahlers:2009kh}.
Strong constraints on WISPs often arise from these effects~ 
\cite{Redondo:2008en}, but they are certainly
model dependent~\cite{Masso:2005ym,Masso:2006gc,Mohapatra:2006pv,Jaeckel:2006xm,Brax:2007ak}
and they rely on our very imperfect knowledge of cosmology, stellar  
interiors or astrophysical environments.

In turn, these arguments suggest that the detection of a WISP in a  
laboratory experiment
will deeply change our view of cosmology and astrophysics. In fact,  
neutrinos (the only
SM WISP) were postulated and confirmed as subtle effects in laboratory  
experiments and nowadays
they are essential ingredients in our understanding of cosmology (like  
in big bang nucleosynthesis
or structure formation) and astrophysics (like in white dwarf cooling  
or supernova type-II explosions).
Furthermore, WISPs have been recently shown to have interesting  
technological applications~\cite{Stancil:2007yk,Jaeckel:2009wm}.

We find that laboratory experiments \emph{at the low energy frontier}~ 
\cite{WP09}
are therefore complementary to collider efforts in the search of an  
accurate description of nature.

A very spectacular effect of such WISPs takes place when we have a photon-WISP
interaction vertex in our theory, a so-called mixing term. 
In this context, photons can convert into WISPs during their propagation, and the
quantum amplitudes of these transitions at different points along the trajectory can 
add up coherently, enhancing the signal, or decreasing it if they add up out of phase. 
This phenomenon is known as quantum flavor oscillations and has been observed in the 
context of the neutral kaon, B~meson and neutrino systems.

\begin{figure}[t]	
\centering
\includegraphics[width=9cm]{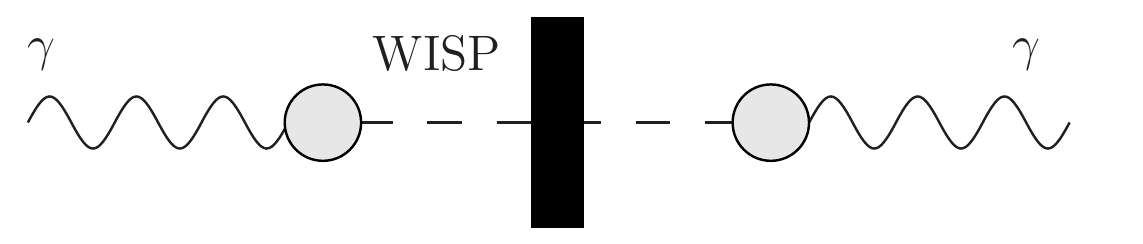}
\vspace{-0.3cm}
\caption{Schematic overview of a light-shining-through-walls experiment. The gray blob indicates the mixing term between photons and the ``weakly interacting sub-eV particle'' (WISP).
\label{fig:lsw} }
\vspace{-0.3cm}
\end{figure}

There are several direct and indirect experimental approaches for WISP searches~\cite{Adelberger:2009zz,Battesti:2007um,Andriamonje:2007ew,Arik:2008mq,Cantatore:2008ue,Inoue:2008zp,Davoudiasl:2008fy,Duffy:2006aa,Melissinos:2008vn,Dobrich:2009kd,Jaeckel:2007ch,Jaeckel:2008sz,Gninenko:2008pz,Gies:2006ca,Gies:2006hv,Gies:2009wx}.
Possibly, the cleanest way to search for photon-WISP oscillations is through so-called 
``light-shining-through-walls'' (LSW) experiments \cite{Okun:1982xi,Anselm:1986gz,VanBibber:1987rq}. 
In such experiments a laser beam is shone onto a thick wall where photons are stopped but WISPs 
produced in oscillations can traverse.  
By the inverse process, WISPs can re-oscillate into photons behind the 
wall and may be detected in a low background environment, see Fig.~\ref{fig:lsw}

%%%%%%%%%%%%%%%%%%%%%%%
%%%%%%%%%%%%%%%%%%%%%%%

Since the number of photons regenerated after the wall is proportional to the 
flux in the ``oscillation'' region, it is crucial to get the highest possible laser powers. 
In this respect, it is extremely disappointing that photons in the oscillation 
region are lost when hitting the wall. 
To partially overcome this situation, in a pioneer LSW experiment~\cite{Ruoso:1992nx,Cameron:1993mr} a so-called optical delay line was set up which forced the light to travel on almost collinear paths for $\approx 200$ times through the oscillation region, thus
augmenting the effective laser power by a factor $\approx 100$.
This scheme suffers however from two drawbacks. 
The first is the necessity of a detector with a large sensitive area because the different beam paths must not overlap in order to avoid destructive interference. 
However, in general the dark noise of a detector for single photons grows with the size of its sensitive area. Thus it will increase with an increasing number of light passes in the delay line and by this deteriorating the signal to noise ratio of the detector. 
The second drawback is that delay lines with a much higher number of light passes than 200 are difficult to construct due to the growing complexity of the underlying geometrics. 

These problems can be solved by the use of optical resonators. Since they are based on the superimposition of all the different equivalent roundtrips within the area of only one light path, the sensitive area of the detector can be kept very small resulting in low dark noise.  
On the other hand, long optical cavities with equivalent round trip numbers of several thousand are used by some large-scale experiments such that they boost up the power inside to levels of several tens of kW. Good examples of these are the laser interferometric gravitational wave detectors like GEO600~\cite{Luck:2006ug} or LIGO~\cite{The07}.

Due to this power enhancement effect optical cavities could be extremely useful in a LSW experiment where they would act as amplifiers for the luminosity that is available from the laser source~\cite{Hoogeveen:1990vq,Sikivie98}. 
However, such a set-up was never realized so far%
\footnote{Cavities with excellent characteristics have been combined with strong magnetic fields 
in other types of WISP searches, like in the PVLAS~\cite{Zavattini:2005tm,Zavattini:2007ee} or 
Q\&A~\cite{Chen:2006cd} searches of WISP-induced dichroism and birefringence.}. 
The main reason for that is the by far increased complexity of a cavity-enhanced LSW-scheme.  
This complexity arises on the one hand from strict constraints on the experimental ambient conditions such as high cleanliness or large vacuum chambers~\cite{Luck:2006ug,The07,Bat07}. On the other hand, in some important cases, a strong magnetic field in the background of the oscillation region is required for oscillations to happen.

In this paper we report on the first successful realization of such a cavity-enhanced LSW experiment, the
``Any-Light-Particle-Search'' (ALPS) experiment~\cite{Ehret:2007cm,Ehret:2008sj} at DESY Hamburg.
It utilizes a long optical resonator on the experiment's production side inside a superconducting HERA dipole. 
The resonator increases the effective laser power for WISP searches by more than a factor of 40. 
Its suitability as part of a large-scale LSW experiment is demonstrated by a first search for WISPs at ALPS. 
The current limitations for its sensitivity enhancement together with other possible improvements of the experiment are identified. 
This should enable us to increase the sensitivity of the whole experiment significantly in the future.

This paper is organized as follows: In Sec.~\ref{WISPs} we briefly describe the WISPs we can search for in the ALPS experiment 
and characterize the phenomenon of oscillations. In Sec.~\ref{sec:ALPS} we detail the components of the ALPS experiment.
The design and characterization of the optical cavity are described in Sec.~\ref{sec:cavity} while in Sec.~\ref{sec:analysis} 
we explain the process of data taking, review the performance of the optical resonator and the detector during the measurement run and present the results.

%%%%%%%%%%%%%%%%%%%%%%%%%%%%%%%%%%%%%%%%%%%%%%%%%
%%%%%%%%%%%%%%%%%%%%%%%%%%%%%%%%%%%%%%%%%%%%%%%%%
%%%%%%%%%%%%%%%%%%%%%%%%%%%%%%%%%%%%%%%%%%%%%%%%%
\section{Photon conversion and the WISP Zoo}\label{WISPs}
%%%%%%%%%%%%%%%%%%%%%%%%%%%%%%%%%%%%%%%%%%%%%%%%%
%%%%%%%%%%%%%%%%%%%%%%%%%%%%%%%%%%%%%%%%%%%%%%%%%
%%%%%%%%%%%%%%%%%%%%%%%%%%%%%%%%%%%%%%%%%%%%%%%%%
%
%
The equations of motion of the photon-WISP system as a function of length can be written as
\begin{equation}
\label{eom}
i\frac{d}{d L} \left( \begin{array}{c} |\gamma\rangle \\ |\phi\rangle \end{array}\right) = \frac{1}{2\omega}
\left( \begin{array}{cc} -2\omega^2 \Delta n & \delta \\
                                       \delta                &  m_\phi^2  \end{array}\right)
\left( \begin{array}{c} |\gamma\rangle \\ |\phi\rangle \end{array}\right)
\end{equation}
where $\omega$ is the photon energy, $\Delta n=n-1$ with $n$ the photon refraction index in the medium, 
$m_\phi$ is the WISP mass and $\delta$ is the quantum mechanical amplitude of the 
 $\gamma\to \phi$ forward transition $\delta = \langle \phi | {\cal H}_{\rm int} | \gamma \rangle$.

The transition probability shows the characteristic oscillatory behavior as a function of the distance $L$
\begin{equation}
\label{prob}
P(\gamma\to \phi) = \frac{4 \delta^2}{M^4}\sin^2 \frac{M^2 L}{4\omega}
\end{equation}
with $M^2=((2\omega^2\Delta n+m_\phi^2)^2+4\delta^2)^{1/2}$. 
Unfortunately, for visible light one usually has $\Delta n>0$ so one cannot match $2\omega^2\Delta n+m_\phi^2=0$
by choosing a suitable medium to maximize the amplitude of the oscillations.
The maximum amplitude is therefore obtained \emph{in vacuum} where $\Delta n =0$ and 
$M^2=(m_\phi^4+4\delta^2)^{1/2}$.
In coherent conditions $M^2 L/4\omega \ll \pi/2$ the probability takes a simpler form
\begin{equation}
\label{prob_simple}
P(\gamma\to \phi) = \delta^2 L^2/(4\omega^2)
\end{equation}
where the coherent enhancement in the interaction length $L$ is evident.

Lorentz invariance forbids $\gamma\to\phi$ transitions when the WISP spin is different from 1, 
and has to be explicitly broken if we want to prove oscillations into spin-0 (or $>1$) particles.
In the ALPS setup this is done by the inclusion of a strong magnetic field $B^{\rm ext}$ 
orthogonal to the propagation direction.
In this case, photons polarized along the magnetic field or perpendicular to it can (and will) behave differently. They will have different $\Delta n$ and $\delta$
so their WISP oscillation probability will be generically different. 

The first WISPs to consider are axions and axion-like-particles (ALPs).
They are well motivated spin-0 particles that couple to two photons depending
on their intrinsic parity via interaction terms in the Lagrangian such as
\begin{equation}
{\cal L}^-_{\rm int} = g_- \phi E\cdot B \quad {\rm and/or } \quad {\cal L}^+_{\rm int}=g_+ \phi\frac{1}{2}(E^2-B^2) 
\end{equation}
where $g_\pm$ are dimensionful coupling constants, $\phi$ the ALP field and $E,B$ the electric and magnetic fields. 
In models where parity is non-conserved~\cite{Hill:1988bu} both couplings are allowed. Such \emph{schizons} 
have very similar phenomenology to the ALPs studied here~\cite{Liao:2007nu,Redondo:2008tq}.

Parity odd ALPs ($0^-$) arise as Nambu-Goldstone bosons
\cite{Nambu:1961tp,Goldstone:1962es,Weinberg:1972fn} (GB)
of spontaneously broken global axial symmetries, 
being the QCD axion~\cite{Weinberg:1977ma,Wilczek:1977pj,Peccei:1977hh,Peccei:1977ur,Dine:1981rt,Zhitnitsky:1980tq,Kim:1979if,Shifman:1979if,Krauss:1986wx,Peccei:1986pn}
the most famous but not unique example~\cite{Wilczek:1982rv,Chikashige:1980qk,Chikashige:1980ui, Chun:1991xm,Nelson:1993nf,Coriano:2005js,Coriano:2006xh,Svrcek:2006yi}.
Parity even ALPs ($0^+$) can be GBs, but also quintessence fields~\cite{Wetterich:1987fm,Peebles:1987ek,Frieman:1995pm,Khoury:2003aq} or fields governing the sizes of extra dimensions (moduli) or gauge couplings (dilatons) in string theories.
Such particles are in principle subject to strong constraints from deviations of Newton's law~\cite{Dupays:2006dp}, although some models which evade astrophysics also overcome these problems~\cite{Masso:2006gc,Brax:2007ak}

Further WISP candidates are hidden photons (HPs)~\cite{Okun:1982xi}.
They are gauge bosons of U(1)$_{\rm h}$ symmetries of a hidden sector, which are ubiquitous in extensions of the 
standard model, e.g.~\cite{Fayet:1980ss,Fayet:1980rr,Barbieri:1986sj,Foot:2000vy}.
Gauge invariance allows HPs to have a St\"uckelberg mass~\cite{Stuckelberg:1937,Stuckelberg:1938} 
and kinetic mixing with the standard photon~\cite{Holdom:1985ag,Foot:1991kb}
\begin{equation}
{\cal L}_{\rm mix} = \frac{1}{2}\chi F_{\mu\nu}B^{\mu\nu}\ , 
\end{equation}
with $F^{\mu\nu}(B^{\mu\nu})$ the photon (HP) field strength tensors and $\chi$ a dimensionless coupling
ranging typically from $\sim10^{-23}$ to $\sim10^{-2}$~\cite{Dienes:1996zr,Abel:2003ue,Abel:2006qt,Abel:2008ai}.
Models with kinetic mixing have a wide range of phenomenological applications, 
e.g.~\cite{Feldman:2007wj,Pospelov:2007mp,Redondo:2008ec,Ibarra:2008kn,Chen:2008yi,ArkaniHamed:2008qn,Feldman:2008xs,Ibarra:2009bm,Batell:2009yf,Rahman:2009rq}.

As a final example, hidden sector particles charged under U(1)$_{\rm h}$, get a non-zero electric charge 
through the kinetic mixing term~\cite{Holdom:1985ag}
\begin{equation}
Q_{\rm MCP} = \frac{e_{\rm h}}{e}\chi\ , 
\end{equation}
where $e_{\rm h}$ is the gauge coupling of U(1)$_{\rm h}$ and $e$ the electron's charge. 
Other MCP model beyond this paradigm can be found in~\cite{Batell:2005wa}.
Since $\chi$ is usually very small we call them mini-charged particles (MCPs).
Photons propagating in external magnetic fields will pair produce MCPs~\cite{Gies:2006ca,Ahlers:2006iz}
and MCP loops can produce photon$\to$HP oscillations even if the HP mass is zero~\cite{Ahlers:2007rd}.

\begin{figure*}[t]	
\centering
\includegraphics[width=16cm]{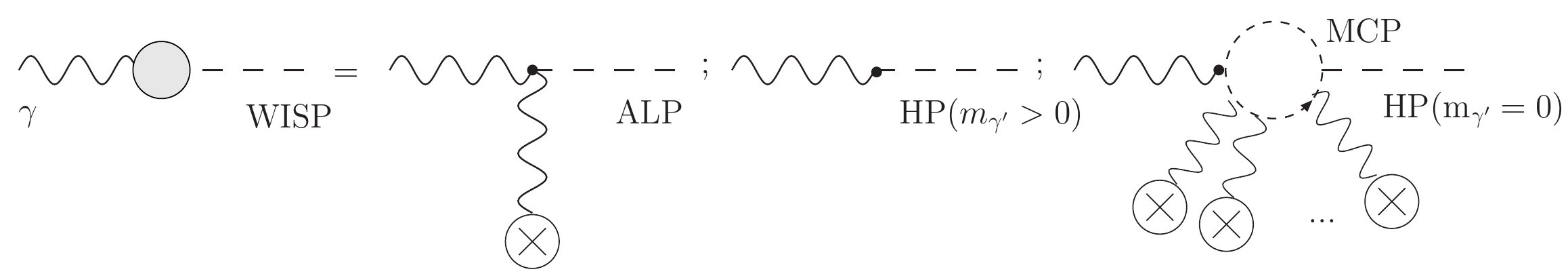}	
\vspace{0cm}
\caption{Feynman diagrams responsible for the mixing term between photons and different hypothetical ``weakly interacting sub-eV particles'' (WISPs). 
Photon oscillations into Axion-like particles (ALPs) and massless hidden photons (HPs) via mini-charged particle (MCP) loops require the presence of a background magnetic field, denoted by a crossed circle. 
Oscillations into massive Hidden Photons occur regardless of such a background. 
See Table~\ref{tab:WISPstuff} for details on the photon-WISP couplings and references.
\label{fig:blobs} }
\vspace{0.45cm}
\end{figure*}

The Feynman diagrams giving rise to photon-WISP mixing in these different models are shown in Fig.~\ref{fig:blobs} and the parameters characterizing the oscillation amplitudes in Tab.~\ref{tab:WISPstuff}.
%%%%%%%%%%%%%%%%%%%%%%%
%
\begin{table*}[t]
\centering
\begin{tabular}{cccccc}
\hline
WISP          & Needs $B^{\rm ext}$ & $\delta_{||}$                             &  $\delta_\perp$                             & $m_\phi^2$          & Ref.       \\[1pt] \hline
Parity odd ALP ($0^-$)   &  yes                                  & $g_- B^{\rm ext}\omega$        &  0                                                   & $m_{\phi_-}^2$    & \cite{Raffelt:1987im}  \\   
Parity even ALP ($0^+$)  & yes & 0                                              &  $g_+ B^{\rm ext}\omega$            & $m_{\phi_+}^2$  &  \cite{Biggio:2006im}   \\
HP ($\muu > 0$)             & no   & $\chi \muu^2$                          &  $\chi \muu^2$                               & $\muu^2$           & \cite{Okun:1982xi}     \\
MCP+HP (for $\muu = 0$)    & yes & $-2\chi \omega^2 \Delta N_{||}$ & $-2\chi \omega^2 \Delta N_\perp$  & $-2\omega^2 \Delta N_{||,\perp}$ & \cite{Ahlers:2007rd}     \\
\hline
\end{tabular}
\vspace{0.2cm}
\caption{Parameters characterizing the photon-WISP probability of oscillations eq.~\eqref{prob_simple}. 
The mixing parameter $\delta$ depends on the relative orientation of the photon polarization and the external field which can be parallel $\delta_{||}$ and perpendicular $\delta_\perp$.
Note that in the MCP+HP case, the indices of refraction due to MCP loops $\Delta N_{||,\perp}=\Delta N_{||,\perp}(Q_{\rm MCP},B^{\rm ext})$ are generally complex and eq.~\eqref{prob_simple} does not hold. See~\cite{Ahlers:2007rd} for the adequate expression.}
\label{tab:WISPstuff}
\end{table*}
\newpage

%%%%%%%%%%%%%%%%%%%%%%%%%%%%%%%%%%%%%%%%%%%%%%%%%
%%%%%%%%%%%%%%%%%%%%%%%%%%%%%%%%%%%%%%%%%%%%%%%%%
%%%%%%%%%%%%%%%%%%%%%%%%%%%%%%%%%%%%%%%%%%%%%%%%%
\section{The ALPS experiment}\label{sec:ALPS}
%%%%%%%%%%%%%%%%%%%%%%%%%%%%%%%%%%%%%%%%%%%%%%%%%
%%%%%%%%%%%%%%%%%%%%%%%%%%%%%%%%%%%%%%%%%%%%%%%%%
%%%%%%%%%%%%%%%%%%%%%%%%%%%%%%%%%%%%%%%%%%%%%%%%%
%%%%%%%%%%%%%%%%%%%%%%%

\begin{figure*}[th]	
\centering
\includegraphics[width=16.13cm]{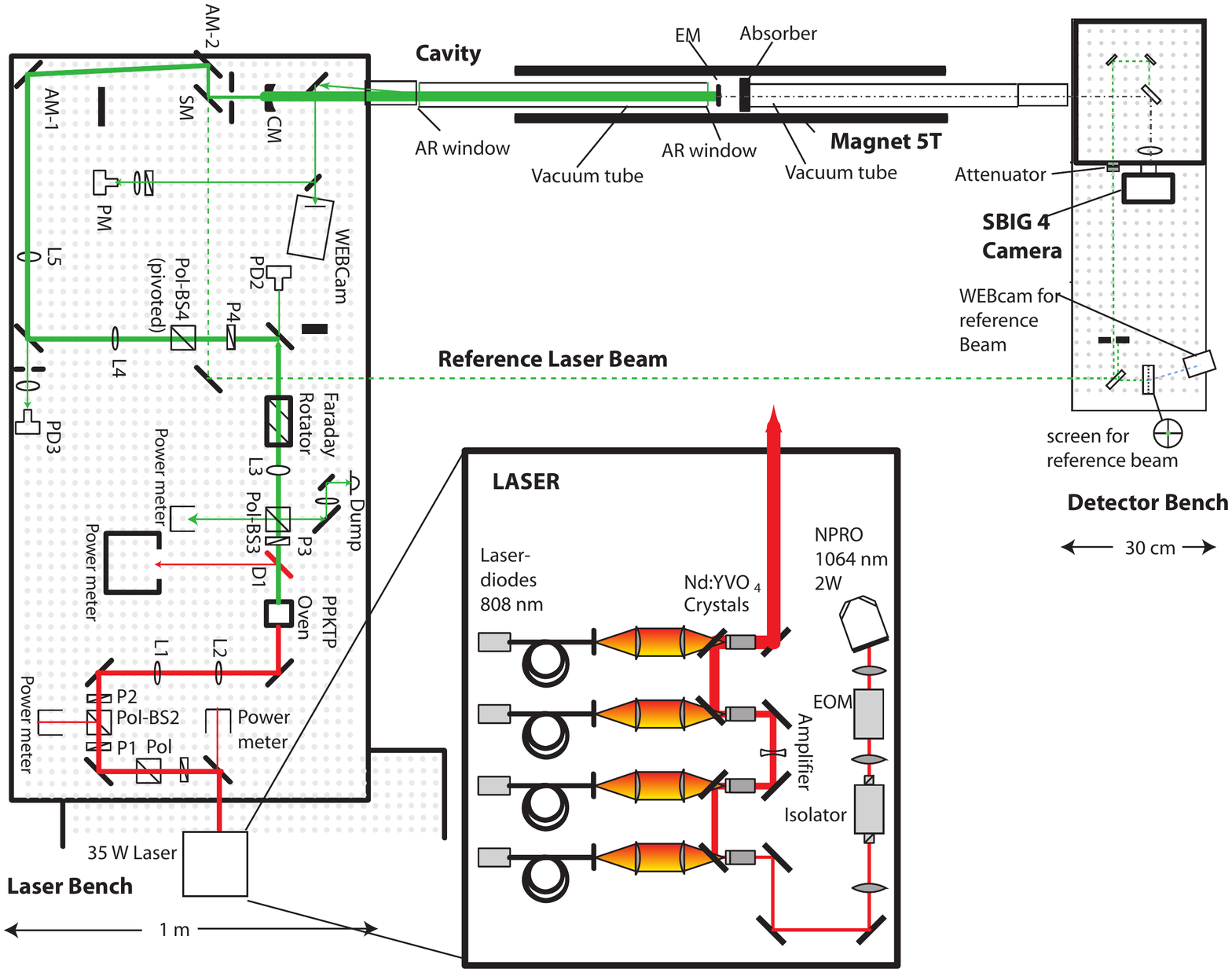}	
\caption{
Schematic overview of the whole experimental setup of the ALPS experiment comprising laser, second harmonic generation in the PPKTP crystal, production cavity, magnet and detector. Magnified is the schematic representation of the ALPS master-oscillator power amplifier laser system. 
\label{fig:setup}}
\end{figure*}

The ALPS experiment as sketched in Fig.~\ref{fig:setup} is built up along a superconducting HERA dipole magnet which provides a field intensity of $B=5.30$~T in a length of 8.8~m. 
Two vacuum tubes are inserted into the magnet at each of both ends, featuring 6.3 and 7.6~m length, respectively.
WISP production takes place in the first tube; regeneration of photons in the second one. 
Around the first tube we have arranged an optical resonator to increase the power available for 
WISP production. A light-tight plug is placed at the inside end of the second tube as our ``wall'' to absorb photons
leaving the cavity towards the regeneration tube.

%%%%%%%%%%%%%%%%%%%%%%%
%%%%%%%%%%%%%%%%%%%%%%%

As the light source a continuous-wave emitting laser at 1064~nm wavelength is used. For detection reasons this light is frequency doubled with a second harmonic generator before it enters the optical resonator that constitutes the production part. Regenerated photons are re-directed by an oblique mirror and focused by a lens into a few pixels of our CCD camera. The volume containing the hypothetical regenerated beam is sealed against environmental light entering into the signal region of the camera. Right at the entrance of the resonator a very small part is separated from the incident beam as a reference for the alignment of the beam axis of the resonator. This reference beam is guided along the magnet's side from the laser bench to the detection bench. 
After very strong attenuation the reference beam is also focused by the same lens onto a different position of the CCD.

In the remainder of this section we present the most relevant characteristics of the different components of our setup.

\subsection{Laser}\label{sec:laser}
The beam tube inside the HERA dipole magnet is bent horizontally and leaves an aperture of only 14~mm with the vacuum tubes installed. This strongly constrains the beam quality of a suitable laser~\cite{Ehret:2007cm}. Furthermore, to efficiently increase the optical power with a cavity, one needs a continuous-wave laser that emits a single longitudinal and transversal mode with a much smaller linewidth than the one of the resonator. Thus the laser source used for the ALPS experiment is a narrow-linewidth master-oscillator power amplifier system (MOPA) operating at 
\unit[1064]{nm} (Fig.~\ref{fig:setup}).
It is based on the system~\cite{Fre07} developed for gravitational wave detectors like LIGO, GEO and VIRGO.

Stable narrow-linewidth emission is provided by a non-planar ring oscillator (NPRO) \cite{Kan85}, emitting \unit[2]{W} of output power with a spectral width of \unit[1]{kHz} (measured over \unit[100]{ms}) and a long term frequency stability of \unit[1]{MHz/min}. In four diode pumped Nd:YVO$_4$ amplifier stages the output power of this laser is increased to \unit[35]{W} preserving the spectral emission characteristics and the nearly diffraction limited beam quality with a fundamental transverse mode content of 95\%. 
The MOPA is equipped with several frequency control elements, which are applied for the cavity frequency locking scheme in the ALPS experiment. A piezo-electric transducer installed on the NPRO laser crystal allows for a frequency shift of $\pm$\unit[100]{MHz} with a response bandwidth of \unit[100]{kHz}. Slow frequency drifts can be compensated by controlling the crystal temperature with a tuning coefficient of \unit[-3]{GHz/K}. Before amplification the NPRO beam is passed through an electro-optic modulator (EOM).

Outside the laser box, the amplifier power can be monitored with a photo-detector (PD1) placed behind a highly reflective mirror. The amplifier output beam has a polarization extinction ratio of more than \unit[20]{dB}. 
A polarizing beam splitter (Pol-BS1) is used to filter out further possible residual light in unwanted polarization states.

\subsection{Second harmonic generation}\label{sec:SHG}
Our experiment uses a detector with a silicon CCD chip whose sensitivity is strongly peaked around the visible spectral region while it approaches zero for wavelengths above 1000~nm~\cite{sbigWWW}. In the phenomenon of oscillations the regenerated light has the same characteristics as the laser beam in the WISP production vacuum tube~\cite{Adler:2008gk}. Therefore we convert the infrared laser light from \unit[1064]{nm} to green \unit[532]{nm} light exploiting the nonlinear effect of second harmonic generation~\cite{Boy68,Spi04} (SHG).

As the nonlinear material we use PPKTP (periodically poled  KTiOPO$_4$) which shows high conversion efficiencies due to its high intrinsic non-linearity. It is fabricated from a flux-grown KTP crystal by reorientation of its optical axis in periods of approximately $\unit[9]{\mu m}$ length by strong electrical poling leading to a much greater conversion efficiency and an easier phasematching condition than for conventional SHG crystals~\cite{Spi04}. The crystal's dimensions are $\unit[1]{mm}\times\unit[2]{mm}\times\unit[2]{cm}$ and its nonlinearity was measured to be $d_{\rm eff}\approx 7.9\cdot 10^{-12}$ m/V. 
It is placed inside an oven to stabilize its temperature at around \unit[38]{°C} in order to maintain phasematching of the infrared and green (SHG) waves. Two lenses (L1, L2, see Fig.~\ref{fig:setup}) are used to focus the infrared beam into the crystal to a waist size of $\unit[135]{\mu m}$. This waist size is a compromise between high conversion efficiency, the risk of damaging the crystal and degradation of the green beam shape. The input polarization can be adjusted to maximum conversion efficiency by a $\lambda/2$-plate (P2). The input power level of the infrared beam is set via a variable attenuator consisting of another $\lambda/2$-plate (P1) and a polarizing beam splitter (Pol-BS2). Behind the oven the converted green light is separated from the infrared by means of a dichroic mirror (D1) which is followed by a variable attenuator for the green light beam (P3, Pol-BS3) and a collimating lens (L3). Then the light passes a Faraday rotator that forms an optical diode together with the polarizing beamsplitter Pol-BS3 to protect the SHG from any back-reflection from the following optical setup.

With this single-pass SHG scheme a long-term stable output power of \unit[800]{mW} at \unit[532]{nm} is available behind D1 when the infrared beam is set to its maximum power level of \unit[35]{W}.

In steady-state the transversal shape of the green beam acquires a slightly doughnut-like form, probably due to nonlinear absorption processes inside the crystal at high intensities~\cite{Wan04}. This slight deviation of the green beam shape from a pure Gaussian lowers the coupling efficiency to the optical resonator to approximately $80\%$ but does not influence the LSW experiment in any other way.

To increase the SHG efficiency we plan to change to a resonant second harmonic generation. This should increase the available green light power by approximately an order of magnitude.

\subsection{Magnet}\label{sec:Magnet}

The production efficiency of some types of WISPs depends strongly on the strength and length of the background magnetic field (Sec.~\ref{WISPs}). The ALPS experiment at DESY uses a spare HERA dipole magnet~\cite{SAM}. More than 400 of these magnets are incorporated in the HERA proton storage ring, which operated from 1991 to end of June 2007.

A HERA dipole magnet has an outer length of 9.8~m and is operated at cryogenic temperatures of 4.2~K. At ALPS we operate the dipole at a 5.30~T magnetic field. Apart from a few quenches we did not encounter any difficulties. 

The bent magnet bore is shielded by an anti-cryostat allowing to perform experiments inside the magnet at room temperature. 
The temperature is stabilized by flushing Nitrogen at a constant rate and temperature
through the bore. As mentioned already in Sec.~\ref{sec:laser} the small clear aperture imposes special requirements on the laser's beam quality.

The direction of the magnetic field is vertical. 
Henceforth, the polarization of the laser light inside the cavity will be measured with respect to this direction.

\subsection{Detector and detection bench}\label{sec:Detector}
%%%%%%%%%%%%%%%%%%%%%%%%%%%%
\begin{figure}[t]	
\centering
\includegraphics[width=8.5cm,angle=0]{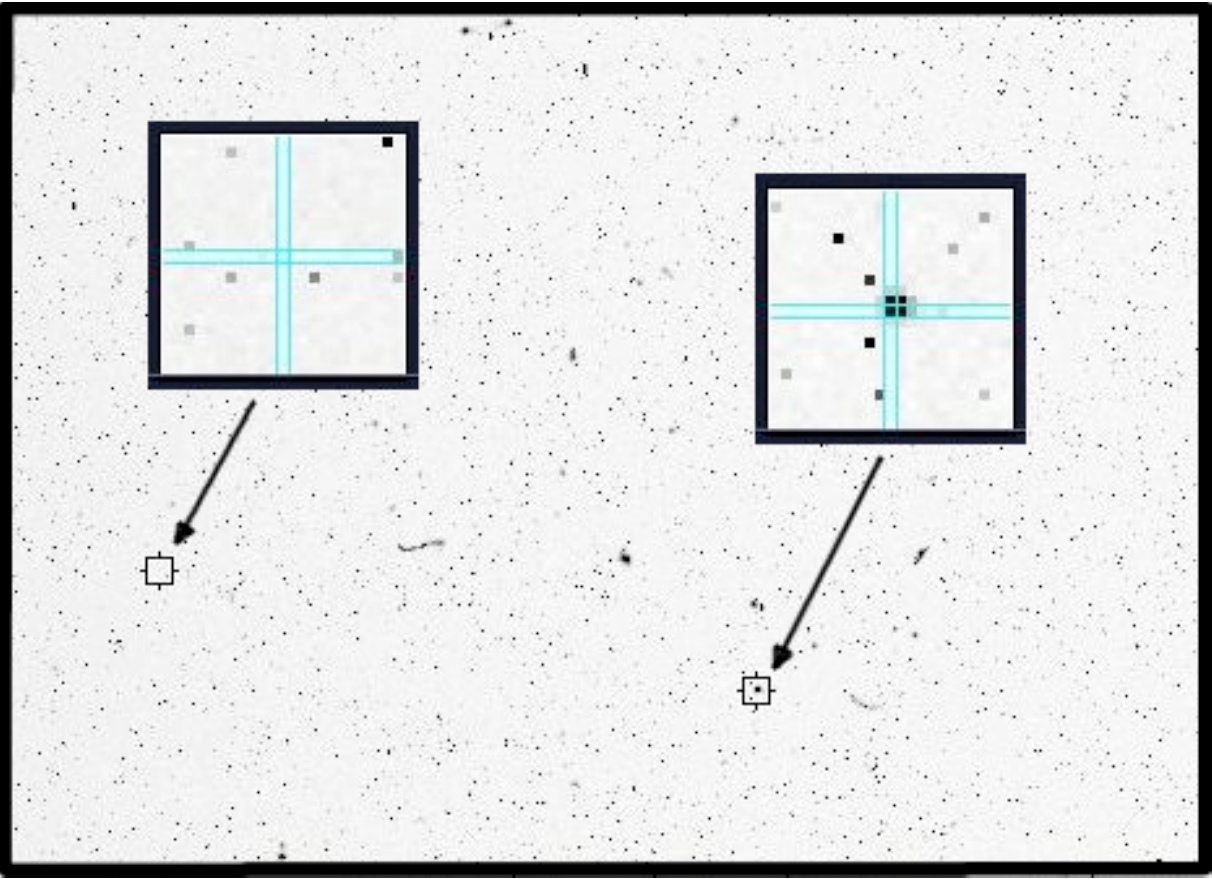}	
\caption{Selection of a typical one hour exposure recorded with the SBIG CCD. 
Tracks from radioactivity and cosmics are visible as well as 
warm pixels, i.e. pixels showing large pedestals or high noise rates. 
The signal (left) and reference (right) regions are shown enlarged. No evidence for photons from WISPs is seen while the reference beam shows up clearly and well focused.}
\label{fig:signal}
\vspace{-0.3cm}
\end{figure}
%%%%%%%%%%%%%%%%%%%%%%%%%%%%

The ALPS experiment uses a commercial astronomy CCD camera SBIG ST-402ME~\cite{sbigWWW} as its detector for regenerated photons. It contains a Kodak KAF-0402ME silicon CCD chip with $765 \times 510$ pixels of 9$\times 9~\mu$m$^2$.
The quantum efficiency is $\approx 60\%$ for light with a wavelength of 532~nm. 

We operate the CCD chip at a temperature of \nobreak{\unit[-\nolinebreak 5]{\textdegree C}} where it shows a low dark current of 0.03 electrons per pixel and second and a readout noise of 17 electrons per pixel, independent of the exposure time. The pixel brightness is measured in ADUs (analog-to-digital units) with one ADU corresponding to 1.46 electrons.

The camera allows sampling times between 0.04~s and 1~h. In order to lower the impact of the  readout noise we used 1~h expositions in our search for WISPs.

In dark frames (exposures with closed shutter) the values of all pixels are found to fluctuate coherently. 
This is monitored by determining the average pixel ADU for each exposure after cutting against cosmics and other spurious signals. This average fluctuates by up to 30 ADUs and is probably caused by some environmental effects. 
To compensate for this effect, the ADUs of all pixels in an individual frame are shifted by the difference of the average pixel value in this frame and an arbitrarily defined baseline. 
After this correction no hint for any additional systematic effect or unknown correlation among the pixels is found.

Special care is required when mounting and fixing all optical components on the detection bench because it has to be removed when the detector tube is pulled out or put into the magnet (as required by the data taking procedure, see Sec.~\ref{sec:analysis}). The exact position of the bench on the supporting structure (fixed to the concrete of the experimental hall) is defined by special feet. In numerous tests it was proved that the focal points of reference beam and alignment beam do not change by more than one pixel on the CCD after removing and re-installing the detector bench.
 
Re-converted photons are reflected by the 45° main mirror and focused by a lens on a few pixels of the CCD in order to keep backgrounds as low as possible.

In order to test the pointing stability a reference beam is split off the main beam in front of the cavity's input coupler (CM) and is guided from the laser bench along the side of the magnet to the detector bench.
A small fraction of this reference beam is reflected by the 45° window to the detector box, attenuated and reflected by two mirrors, so that it passes from behind through the main mirror and is focused on the CCD at a different location than the focus of the main beam.
A typical CCD frame with the two regions of interest is shown in Fig.~\ref{fig:signal}.

In future experiments we plan to improve the signal to noise ratio by one order of magnitude by 
using a more sensitive camera and by reducing the signal spot size.

%%%%%%%%%%%%%%%%%%%%%%%%%%%%%%%%%%%%%%%%%%%%%%%%%
%%%%%%%%%%%%%%%%%%%%%%%%%%%%%%%%%%%%%%%%%%%%%%%%%
%%%%%%%%%%%%%%%%%%%%%%%%%%%%%%%%%%%%%%%%%%%%%%%%%
\section{The ALPS cavity}  \label{sec:cavity}
%%%%%%%%%%%%%%%%%%%%%%%%%%%%%%%%%%%%%%%%%%%%%%%%%
%%%%%%%%%%%%%%%%%%%%%%%%%%%%%%%%%%%%%%%%%%%%%%%%%
%%%%%%%%%%%%%%%%%%%%%%%%%%%%%%%%%%%%%%%%%%%%%%%%%

The purpose of optical resonators (cavities) in the context of LSW experiments is to enlarge the light power in the WISP production region  relative to the available primary laser power. Such an optical resonator acts therefore as an amplifier for the luminosity of the whole experiment boosting its sensitivity.
This is possible by coherent superposition of light fields that enter the resonator at different times corresponding to successive roundtrips. 
The ratio of the laser power inside the resonator traveling towards the regeneration tube, 
$P_{\to}$, to the power incident to the coupling mirror, $P_{\rm CM}$,
is called power build-up $P\!B$. 
\begin{equation}
P\!B =  P_{\to} / P_{\rm CM} \ .
\end{equation}

Consider a linear resonator consisting of two mirrors spaced by a distance $L$ fed by a laser of frequency $\nu$. 
The first mirror or input coupler has a power transmission coefficient of $T_{\rm CM}$ and the second one of $T_{\rm EM}$. 
Absorption, scattering and deflection of the light during one roundtrip are combined into a parasitic loss factor $A$. 

During a roundtrip between the mirrors the light acquires a phase $\psi = 2\pi\nu\, (2 L) / c$. 
Resonant enhancement of the circulating light power is achieved when the light wave nearly reproduces itself after one roundtrip, i.e. when $\Phi = \psi$ mod $2\pi \approx 0$. 
A cavity has therefore an infinite number of resonances at frequencies $\nu_{{\rm res},n}=n c /(2 L)$, characterized by integers $n$. 
The frequency interval between resonances is called \emph{free spectral range}, $F\!S\!R=c/(2 L)$. 

Under the assumption that $T_{\rm CM}$, $T_{\rm EM}$, $A$ and $\Phi$ are all small compared to unity one can approximate the power build-up by~\cite{Sie86}
\begin{equation}
P\!B \approx \frac{4\,T_{\rm CM}}{\left(T_{\rm CM}+T_{\rm EM}+A\right)^2+4\,\Phi^2}\quad , \label{eqn:PBgen}
\end{equation} 
which in resonance gives 
\begin{equation}
P\!B_{max} \approx \frac{4\,T_{\rm CM}}{\left(T_{\rm CM}+T_{\rm EM}+A\right)^2} \quad .\label{eqn:PB}
\end{equation}
The errors of these approximations are negligible for all considerations that are made throughout this paper.

The derivative of eq.~(\ref{eqn:PB}) with respect to $T_{\rm CM}$ can be set to zero to find the maximum power build-up with respect to input coupling.
This maximum is achieved if the so-called impedance matching condition is fulfilled
\begin{equation}
T_{\rm CM}=T_{\rm EM}+A \quad .
\end{equation}
Hence, the largest power build-up is obtained when the transmission of the input coupler $T_{\rm CM}$ is chosen to be as close as possible to the sum of parasitic losses $A$ and output coupling of the cavity $T_{\rm EM}$.

Derivation of eq.\eqref{eqn:PB} with respect to $A$ or $T_{\rm EM}$ shows that there is no maximum of $P\!B_{max}$ over these parameters clarifying that it is always best to keep parasitic losses and the output coupling as small as possible in order to maximize the power build-up. 
But, even when choosing the best optics available and assembling the cavity under as clean ambient conditions as possible, some parasitic losses will always remain. 
Hence, one normally has to guess $A$ in advance during the design process of an optical resonator.
The transmission coefficient of the input coupler can then be chosen to maximize $P\!B_{max}$.

To realize the full possible power build-up in the cavity, the alignment, shape and resonance frequency of the resonator eigenmode must be exactly matched by the incident laser beam. 
These parameters are defined by the alignment and radii of curvature ($R\!O\!C$) of the cavity mirrors and by the optical distance between them~\cite{Kog66}. 

Single mode continuous-wave lasers emit most of their power into the fundamental transversal mode \TEM. 
It is therefore most efficient to use spherical mirrors for the resonator such that the laser mode can be easily matched to the eigenmode of the cavity. 
However, every slight mismatch in beam shape or alignment causes some fraction of the incident power to overlap with higher order transversal eigenmodes of the cavity. 
In general, the higher order spatial modes have different resonance frequencies than the fundamental mode. 
Hence, this power fraction is not coherently enhanced when the fundamental mode is resonant.
This effect is called mismodematching and it has to be minimized in order to maximize the intra-cavity power. 
In turn, such a cavity with non-degenerated resonant frequencies has the positive effect of behaving like a mode filter for the production beam, which also restricts the beam of re-converted photons to exist only in the \TEM mode. 
Such a beam can then be focused to smaller spot sizes than beams consisting of multiple higher order transversal modes, 
which is a key issue when maximizing the signal to background ratio in our CCD camera. 
The full theory on modestructures of laser beams and resonators can be found in several publications, for instance~\cite{Sie86,Kog66,Kwe07}.

Recall that, in addition to an optimized spatial overlap, the laser frequency has to match one resonance frequency $\nu_{\rm res}$.
Small fluctuations of the cavity length $\Delta L$ induce correspondingly small relative changes of its resonance frequency given
by $\Delta \nu_{\rm res}/\nu_{\rm res}=-\Delta L/L$. 
To show how even tiny length changes degrade the power build-up, we solve eq.~\eqref{eqn:PBgen} for an absolute length change $d$ that would cause a reduction to one half of its resonant value, 
\begin{equation}
\frac{P\!B}{P\!B_{max}}\, \overset{!} = \, \frac{1}{2} \quad \Rightarrow \quad d = \pm \frac{1}{8\pi}\lambda\left(T_{\rm CM}+T_{\rm EM}+A\right) 
\label{eqn:d}\ . 
\end{equation}
From this expression (and bearing in mind that we are interested in the smallest possible values of $T_{\rm CM}+T_{\rm EM}+A$ to maximize the power build-up) it becomes clear that even length fluctuations on scales much smaller than the laser wavelength will cause huge changes of the power build-up of the cavity.

A similar constraint for the necessary stability of the laser frequency can be deduced from equation~(\ref{eqn:d}) if one substitutes $\lambda= c / \nu$ and solves for the laser frequency $\nu$ at which the power build-up is reduced to half its peak value. The distance between these two points on the frequency axis is called the \emph{full width at half maximum} (FWHM) linewidth of the optical resonator and is given by
\begin{equation}
\label{eqn:FWHM}
\Delta\nu = F\!S\!R \, \frac{(T_{\rm CM}+T_{\rm EM}+A)}{2\pi} \ .
\end{equation}
The laser frequency must be kept well within this FWHM 
linewidth to keep the power build-up near its peak value.
The ratio $F\!S\!R$/FWHM, which is a measure for the cavity's frequency selectivity, is usually called finesse.

In order to keep the power build-up stable, the difference between the resonance frequency and the laser frequency can be minimized by a feed-back control system. 
This can either actuate on the cavity length or on the laser frequency.

Once spatial modes and frequencies of the laser and the cavity are matched, the cavity will exhibit its full power build-up.
This gain factor will remain fairly constant while scaling the incident power until it gets spoiled by thermal distortion or even by the destruction of the optics.

The power inside the cavity can be easily measured by monitoring the power transmitted through the end mirror, $P_{\rm EM}$ since, under the outlined approximations
\begin{equation}
\label{PEM}
P_{\rm EM}= T_{\rm EM} P_{\to}  \ . 
\end{equation}

\subsection{Design of optical resonator and stabilization}\label{sec:OptRes}

The design of the ALPS optical resonator was constrained by several aspects. 
First, the site at DESY does not allow for the use of two magnets in a row so the production and regeneration regions of our LSW experiment must be located inside the same magnet. 
This requires that one mirror of the cavity has to be placed in the middle of the HERA magnet. 
Second, the bent magnet bore has an cat's pupil-shaped clear inner aperture of only \unit[28]{mm} height and \unit[14]{mm} width (see~\cite{Ehret:2007cm}). 

A linear resonator comprised by two mirrors was realized. 
One mirror, CM, is held in place by a rigid and manually adjustable mirror mount on an optical table in front of the magnet and the other mirror, EM, is mounted near the middle of the magnet on a self-made small and nonmagnetic mirror holder attached to the end of the WISP production vacuum tube (see~\cite{Ehret:2008sj}). 
The mirror holder for EM was designed such that it allows for remote alignment to compensate for alignment drifts.
The distance between both mirrors is \unit[8.6]{m} most of which is occupied by a vacuum tube with two anti-reflectively coated windows (AR windows). 

\begin{figure}[t]	
\centering
\includegraphics[width=6.1cm,angle=-90]{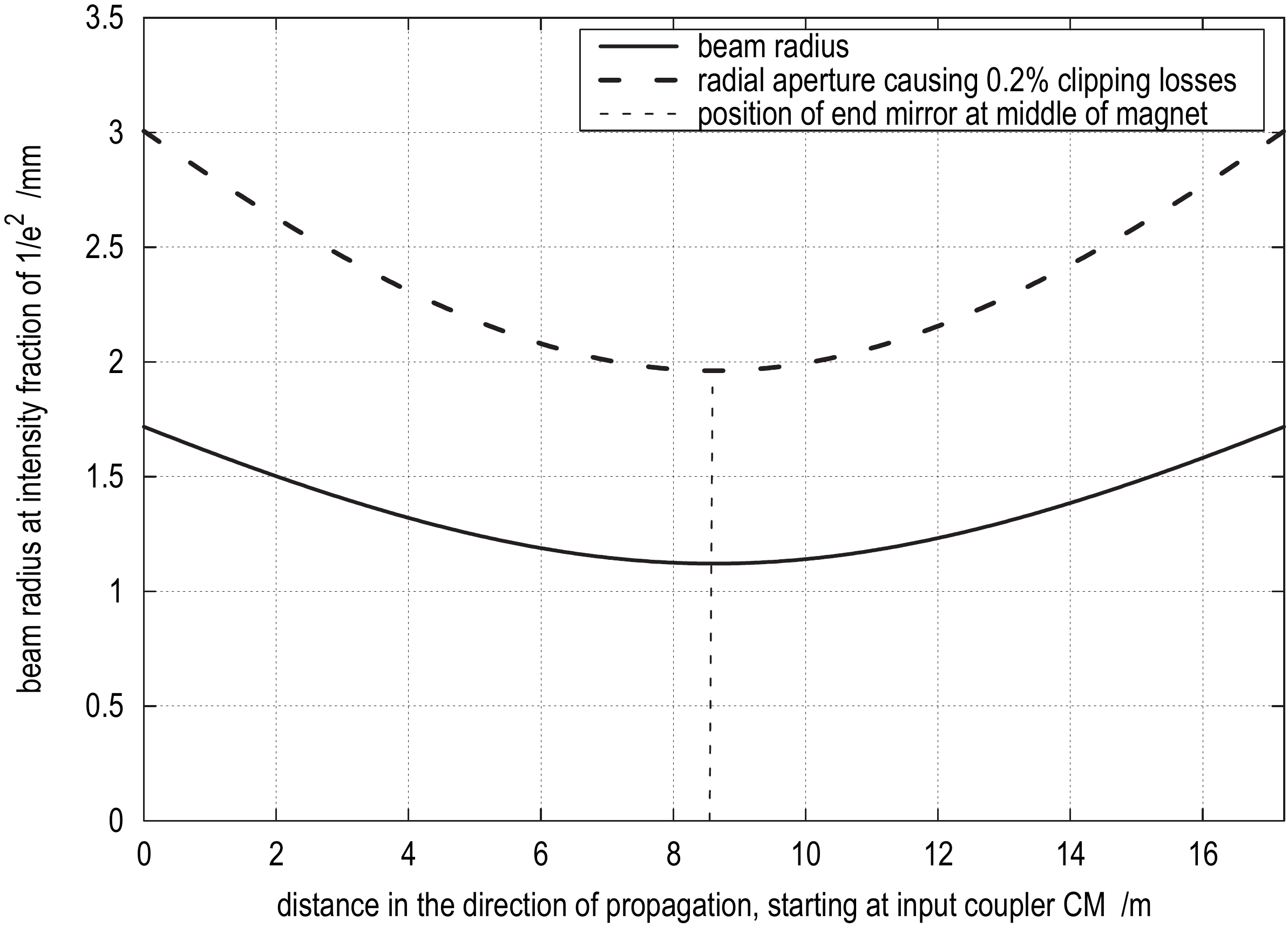}	
\caption{Beam radius of the fundamental eigenmode of the ALPS optical resonator together with the radius at which $\unit[0.2]{\%}$ of the mode's power would be clipped. 
The position of the end mirror EM is shown as a vertical line. 
Clearly, the beam size is always well below the minimum aperture of our production vacuum tube (14 mm).}  
\label{fig:eigenmode}
\vspace{-0.2cm}
\end{figure}

Several stable resonator designs that differ
in the radii of curvature of the cavity mirrors are possible~\cite{Sie86,Kog66}. 
Our resonator uses a plano-concave design with one plane mirror (EM) and a curved one with $ROC=\unit[-15]{m}$. 
The resulting fundamental cavity eigenmode needs a free circular aperture of only \unit[6]{mm} diameter to keep power losses per roundtrip due to clipping below $\unit[0.2]{\%}$. 
The resonance frequencies of all higher order transversal eigenmodes up to a mode index sum of 10 do not match the 
\TEM resonance frequency within a linewidth. 
This simplifies the readout of the length changes of the cavity, which is necessary for the feed-back control system explained below. 
The evolution of the beam radius of the chosen fundamental eigenmode is shown in Fig.~\ref{fig:eigenmode} together with the required free circular aperture to keep losses due to clipping below the value mentioned above. 
Here ``beam radius'' corresponds to the lateral distance from the beam's propagation axis where the Gaussian intensity distribution of the fundamental \TEM has fallen to a fraction of $1/e^2$.

%%%%%%%%%%%%%%%%%%%%%%%%%%%%
%%%%%%%%%%%%%%%%%%%%%%%%%%%%

As end mirror (EM) we chose a highly reflecting mirror whose measured transmission is  $T_{\rm EM} = \unit[170]{ppm}$.
Because our resonator includes four antireflective coated surfaces of the vacuum tube's windows (see Fig.~\ref{fig:setup}), with
rest reflection coefficient of $R\lesssim \unit[0.3]{\%}$ each, we chose a power transmission design value of approximately $\unit[2]{\%}$ for the input coupler. 
Accurate measurements on CM resulted in a power transmission of $T_{\rm CM} \approx \unit[2.3]{\%}$. 
Assuming the realization of an impedance matched cavity according to equations~(\ref{eqn:PB}) and~(\ref{eqn:FWHM}) these values would result in a maximum power build-up of $P\!B_{max}\approx 43$ and a FWHM of $\Delta\!\nu \approx \unit[130]{kHz}$.

In order to get as close as possible to this theoretical value, an optimization of the spatial overlap of laser and resonator mode has to be performed. This can be done with the two beam shaping lenses L4 and L5, the two alignment mirrors AM1 and AM2 and with the two adjustable mounts of the cavity mirrors CM and EM (see Fig.~\ref{fig:setup}). 
If the incident laser beam as well as the cavity mode are perfect Gaussian \TEM modes, then the combination of these actuators must be sufficient to maximize the spatial overlap in all its degrees of freedom.

The experimental site at DESY is exposed to strong acoustic and vibrational influences. These ambient conditions give rise to large and fast length fluctuations of our cavity as well as significant alignment fluctuations with strong impact on the power build-up.
In order to compensate these fluctuations we constructed an electronic feedback control loop that actuates on the laser frequency.

The difference between the laser frequency and the actual resonance frequency of the cavity is determined via a sideband modulation spectroscopy technique called Pound-Drever-Hall (PDH) readout scheme, with the help of a photodiode PD3 whose output is high-pass filtered~\cite{Pou46,Dre83,Bla00}. 
The necessary phase modulation sidebands are generated by an electro-optic modulator (EOM) which is part of the laser system (see Sec.~\ref{sec:laser}). 
After demodulation the resulting error signal is filtered by a controller and then amplified by a fast high voltage amplifier. 
The resulting high voltage signal is then passed to the fast frequency actuator which is part of the laser system (see Sec.~\ref{sec:laser}). 
This control loop has a bandwidth of approximately \unit[28]{kHz} limited by mechanical resonances of the frequency actuator around \unit[200]{kHz}.

The length fluctuations of the cavity are much larger than the range over which the PDH scheme would provide a valid error signal. Therefore, if the control loop is closed at an arbitrary cavity state the controller would in most cases not be able to bring cavity and laser into resonance. To solve this problem the loop is equipped with automatic lock acquisition electronics that monitors the light power that is reflected back from the cavity via the DC signal of PD3. 
If the locked state is lost, the reflected power rises above a user-defined level which is the trigger to scan the laser frequency and search for the next resonance. If this is found the system is automatically relocked. Intervention by an operator is not necessary.

We record the DC signal of PD3 every 30 seconds in
order to keep the information about the intracavity power for the data analysis.
This signal was calibrated in test runs by measuring simultaneously the light passing through the mirror EM with known transmission $T_{\rm EM} \approx \unit[170]{ppm}$.

The polarization state of the laser light entering the cavity can be adjusted by a combination of a $\lambda/2$ plate (P4) and a pivotable polarization beam splitter (Pol-BS4). By this combination a clean linear polarization state with an arbitrarily adjustable angle relative to the magnetic field direction can be realized.

\subsection{Characterization of the production resonator}
\label{sec:cav-characterization}

%%%%%%%%%%%%%%%%%%%%%%%%%%%%%%
\begin{figure}[t]	
\centering
\includegraphics[width=6cm,angle=-90]{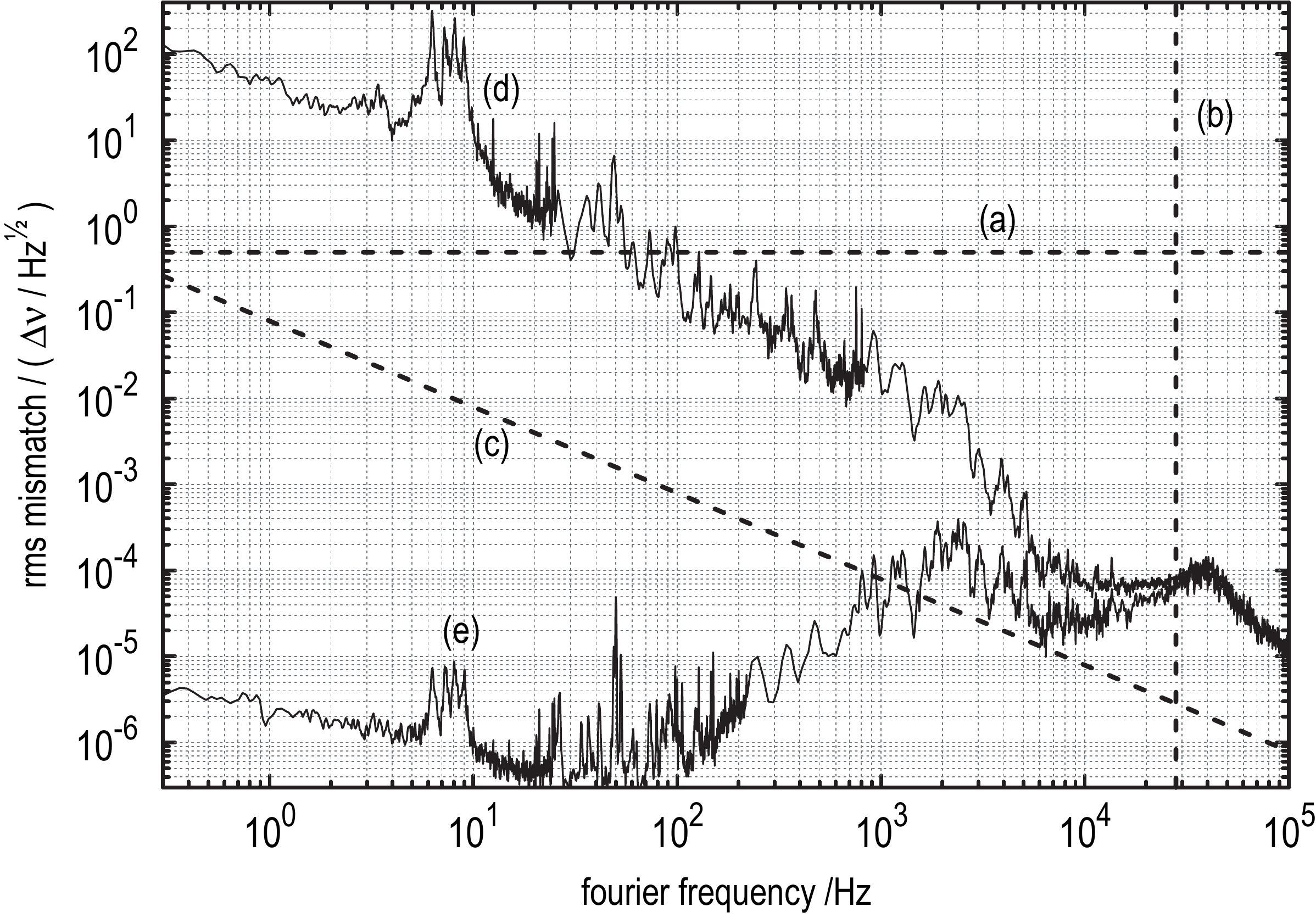}	
\caption{Comparison of rough upper limit fit to free-running laser frequency noise {\bf (c)}, that was taken from~\cite{Heu04}, with measurements of the free-running {\bf (d)} and stabilized {\bf (e)} mismatch between the laser and cavity resonance. The free-running mismatch is effectively suppressed for Fourier frequencies smaller than the bandwidth of the control loop {\bf (b)}. The power build-up is seriously degraded for frequencies with larger rms noise than one half of the cavities FWHM linewidth $\Delta\nu$ {\bf (a)}.}  \label{fig:length noise}
\vspace{-0.3cm}
\end{figure}
%%%%%%%%%%%%%%%%%%%%%%%%%%%%%

After installation the cavity was first characterized at low incident powers $P_{\rm CM}$ around \unit[50]{mW}. The stabilized fundamental eigenmode of the ALPS cavity is observed with a CCD camera behind the mirror EM. It has a nice round shape without any observable fluctuations or distortions. 
This is important for the detection stage of our LSW experiment because the position and intensity distribution of a possible beam of re-converted photons behind the wall is then stationary and well defined by the orientation and profile of the cavity mode (see Sec.~\ref{sec:Detector}). Furthermore, this is a first hint suggesting that the cavity mode does not suffer from large losses due to extensive clipping or heavy absorption.

According to eq.~\eqref{eqn:PB} optical losses inside the cavity would reduce the achievable power build-up. Due to its length of \unit[8.6]{m} and the way it must be mounted inside the magnet, it is not possible to assemble the cavity in a cleanroom environment. Therefore it is important to check if additional internal losses might be caused by dust or other impurities. 

In Sec.~\ref{sec:OptRes} we expected $P\!B_{max}\approx 43$ from our cavity design. 
We measured this quantity by recording the power transmitted through EM and determining the intra-cavity power $P_{\to}$ through 
eq.~\eqref{PEM} in a dedicated setup.
We achieve a power build-up of $P\!B = 44 \pm 2$ which at first glance seems to agree
nicely with our expectations.

However, a fast scan of the laser frequency over more than one free spectral range reveals that $\approx$ \unit[20]{\%} of the incident power $P_{\rm CM}$ is coupled to different higher order transversal modes. 
This value could not be reduced even by careful optimization of the spatial overlap between laser beam and cavity mode. 
Therefore it most likely originates from distortions of the incident beam shape to some non-Gaussian intensity profile by the SHG and from alignment fluctuations of the long cavity~\cite{Kwe07}. 
In order to compare our measurement of $P\!B$ with the value of $P\!B_{max}$ obtained in the previous section
one has to correct for the amount of $P_{\rm CM}$ in higher order transversal modes. This correction for the mismodematching results in a measured power build-up in the \TEM mode of
\begin{equation}
\label{eqn:PTEM00}
P\!B_{_{{\rm TEM}_{00}}}=55 \pm 3
\end{equation}
which is slightly higher than expected. 

Additionally, a value of $\Delta\nu\approx \unit[130]{kHz}$ was expected. 
A direct measurement of this value resulted in $\Delta\nu=\unit[127\pm 12]{kHz}$, 
in excellent agreement with the expectations.

%%%%%%%%%%%%%%%%%%%%%%%%%%%%
\begin{figure}[t]	
\centering
\includegraphics[width=6cm,angle=-90]{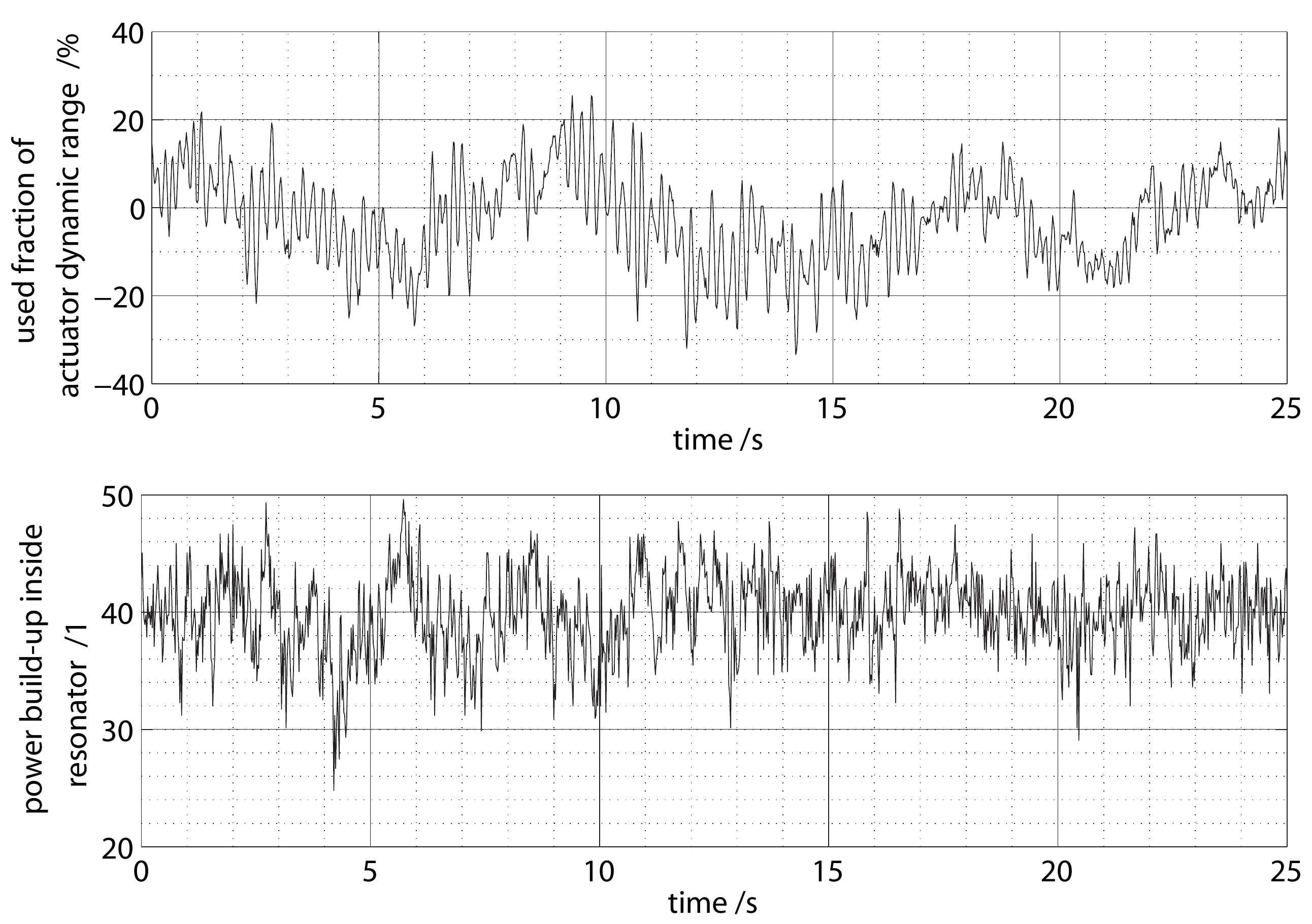}	
% \vspace{-0.5cm}
\caption{
Typical time series of the power build-up (bottom panel) and the actuator signal (top panel) while the electronic frequency stabilization control loop is active.}  
\label{fig:long lock timeseries}
\vspace{-0.3cm}
\end{figure}
%%%%%%%%%%%%%%%%%%%%%%%%%%%%

Therefore, we have two independent ways of determining the internal losses in the cavity. 
Using the measured values of $P\!B_{_{{\rm TEM}_{00}}}$ and $\Delta \nu$ with eqs.~\eqref{eqn:PB} and~\eqref{eqn:FWHM} we obtain for the parasitic losses introduced by the antireflective coated surfaces inside our cavity values of $0.22\pm 0.01\%$ and $0.28\pm 0.05 \%$ per facet, respectively. 
These values are compatible with each other. 
They turn out to be slightly smaller than the estimation used for the design of the resonator, 
which explains the fact that $P\!B_{_{{\rm TEM}_{00}}}$ surpasses our expectations.
Clearly, the windows inside the cavity dominate the overall losses. 
Eliminating the windows in future experiments could be the first step towards a significantly enlarged power build-up.

To characterize the adequacy and performance of our frequency stabilization control loop we further investigate the free-running and stabilized rms fluctuations of the frequency mismatch between laser and cavity. 
The measured spectral densities are shown in Fig.~\ref{fig:length noise}. 
Because the power build-up is seriously degraded if the frequency mismatch amounts to more than half of the cavities FWHM all measurements in Fig.~\ref{fig:length noise} are normalized to its measured value of $\Delta\!\nu=\unit[127]{kHz}$. 
The figure shows the squared sum of the measured noise spectral densities of the control signal applied to the frequency actuator of the laser and of the error signal of the control loop. This gives an estimate on the size of the mismatch in an uncontrolled situation (free-running). 
Additionally, the measured noise spectral density of the error signal of the control loop itself is shown which reflects the remaining mismatch once the control loop is closed.
Furthermore, a line is plotted that gives a rough estimate of the free-running laser frequency noise. 
One can see that the by far greatest contribution to the overall free-running frequency mismatch is due to cavity length noise. 
For Fourier frequencies below \unit[100]{Hz} this free-running rms length noise causes resonance frequency fluctuations much greater than one half of the cavities FWHM linewidth reducing the average power build-up close to zero. 
But when the active stabilization system is turned on the rms frequency mismatch is largely reduced. 
Their remaining impact on the power build-up is virtually negligible.

Our control loop is capable of stabilizing the laser to the cavity on longer timescales as well. 
In Fig.~\ref{fig:long lock timeseries} a typical time series of remaining power build-up fluctuations under lock
is displayed. One can see that the laser light is kept resonant inside the cavity for tens of seconds. Less than $\unit[40]{\%}$ of the dynamic range of the frequency control actuator is needed during this time so that slow drifts over even longer timescales can still be compensated.

The remaining power build-up fluctuations probably originate from changes of the alignment of the incident laser beam relative to the eigenmode of the resonator. 
An additional auto-alignment system may improve this situation and increase the average power build-up slightly. 

As far as the frequency control loop is concerned, very slow drifts of the cavity length saturate the actuator on timescales of some tens of minutes causing the stabilization to fail and hence the power build-up to be lost. 
In that case a fast relock was realized by means of the already mentioned automatic lock acquisition electronics and hence long measurement periods were possible.

%%%%%%%%%%%%%%%%%%%%%%%%%%%%
\begin{figure}[t]	
\centering
\includegraphics[width=6cm,angle=-90]{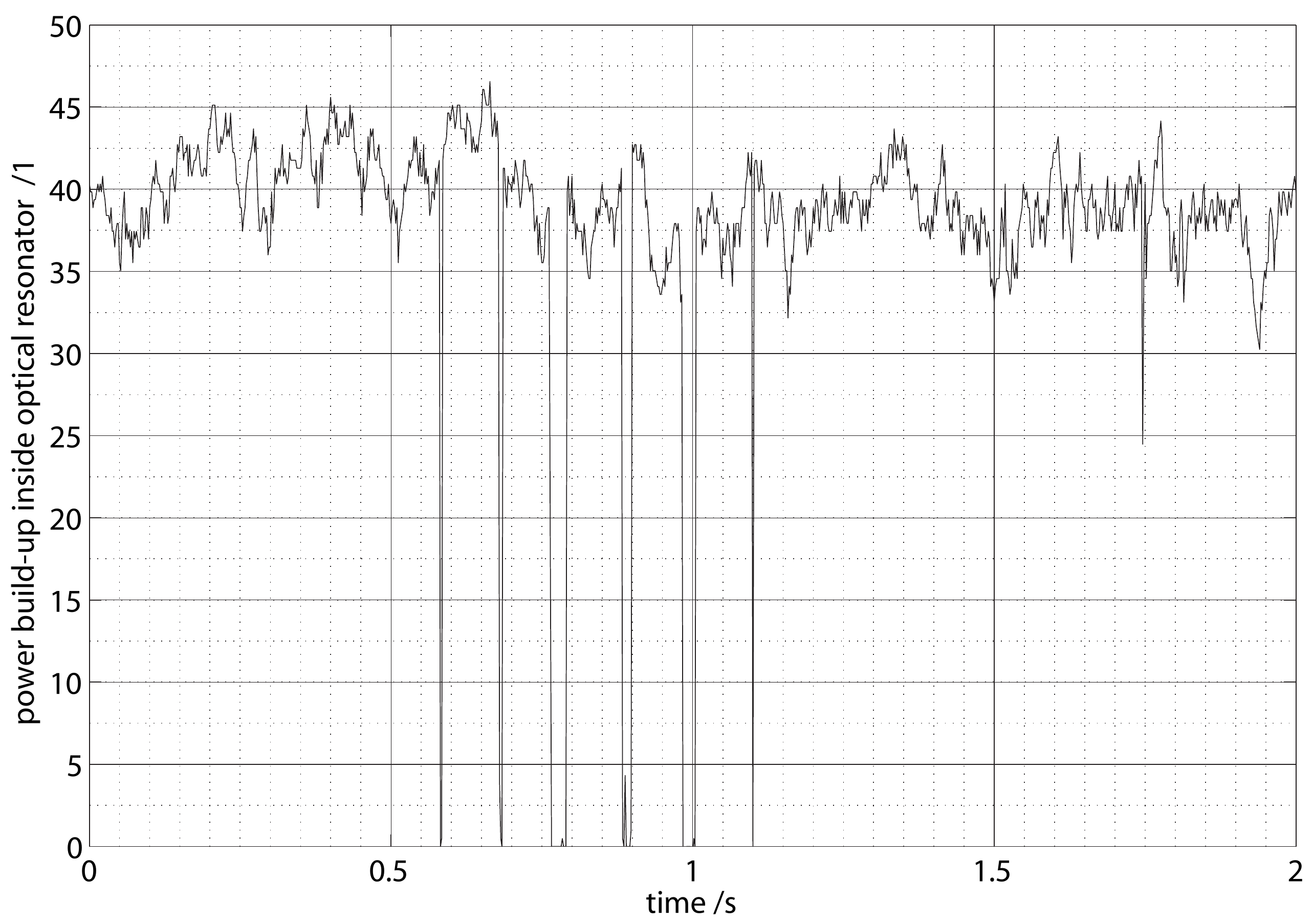}
% \vspace{-0.7cm}
\caption{Typical power build-up inside the optical resonator during multiple automatic relock sequences caused by deliberately introducing a strong vibrational disturbance of the cavity length.}
\label{fig:automatic relock}
\vspace{-0.3cm}
\end{figure}
%%%%%%%%%%%%%%%%%%%%%%%%%%%%

A timeseries of the power build-up during such a relock sequence is shown in Fig.~\ref{fig:automatic relock}. 
To stimulate a loss of the locked state of the resonator strong mechanical length fluctuations were introduced into the system by deliberately hitting the optical table. 
While this disturbance is ringing down the stabilization fails several times because of limited dynamic range which causes the power build-up to drop to zero. 
This is detected by the automatic lock acquisition electronics that reacquire the locked state every time within only fractions of a second.

\begin{figure}[t]
\centering{
\includegraphics[width=0.5\textwidth]{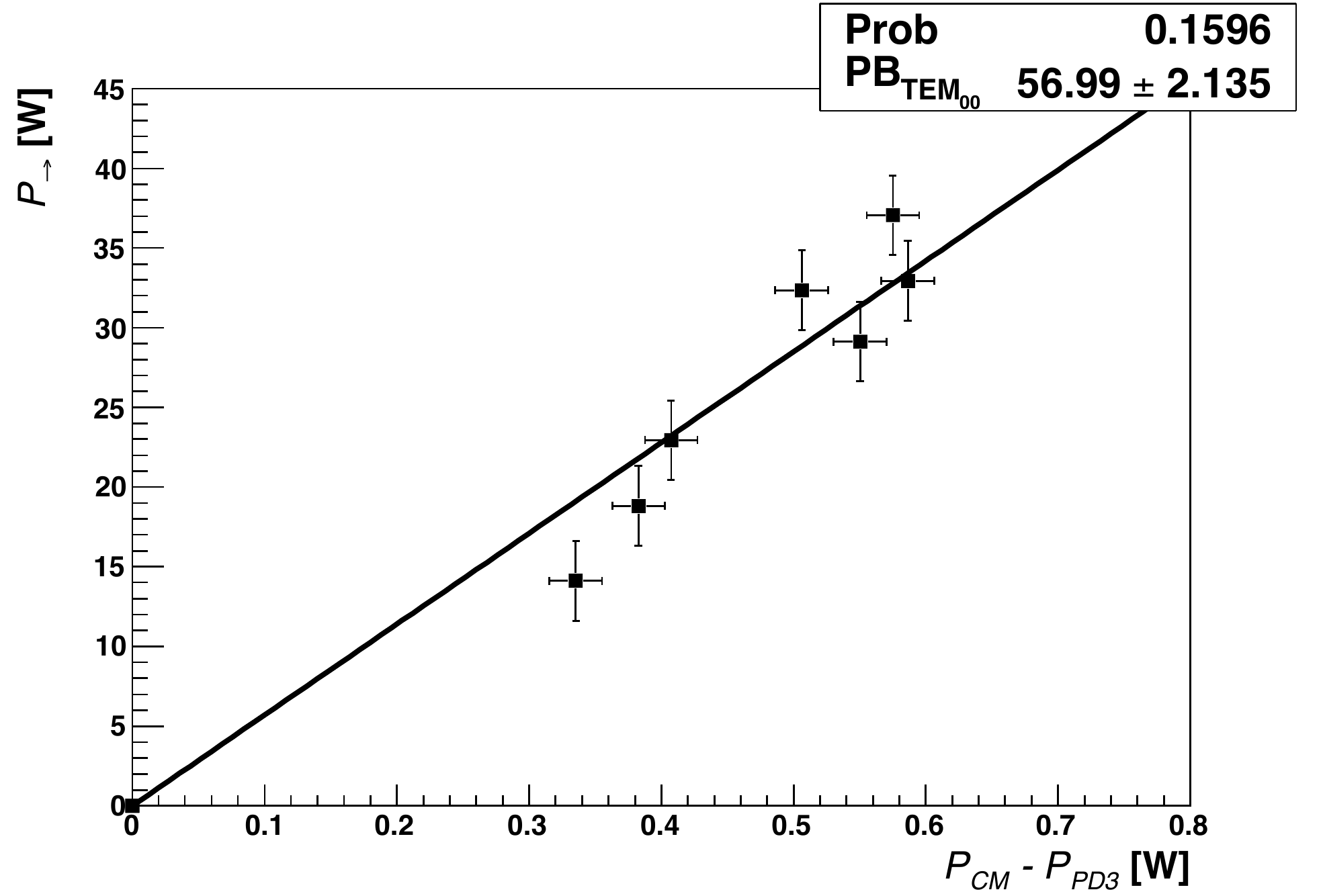}}
\vspace{-0.5cm}
\caption{
Power inside of the cavity, $P_\to=P_{\rm EM}/T_{\rm EM}$, as function of the power fed into the cavity, $P_{\rm CM}- P_{{\rm PD}3}$ where $P_{\rm CM}$ was kept equal to $0.8$ W during the different measurements and $P_{{\rm PD}3}$ denotes the cavity's reflected power as measured by PD3. The slope gives by definition $P\!B_{{\rm TEM}_{00}}$. }
\label{fig:calib}
\vspace{-0.3cm}
\end{figure}

In order to characterize the impact of alignment fluctuations and to correlate the internal cavity power with the power reflected by CM we performed a set of calibration measurements. 
Their results are used to monitor the power in the cavity during data taking.
For the calibration we introduced manually different small misalignments of the laser direction and the cavity's eigenmode and measured the power leaking through the mirror EM and the PD3 signal while the cavity was locked. 
The results are shown in Fig.~\ref{fig:calib}.
They show how the intra-cavity power $P_\to$ is degraded by small misalignments and how this is tracked by the power reflected by CM. 
The slope of this relation gives by definition the power build-up 
$P\!B_{{\rm TEM}_{00}}$. The agreement between this method and the value in eq.~\eqref{eqn:PTEM00} 
demonstrates the overall consistency of our understanding of the cavity.

Finally, let us turn to the characterization of the polarization.
The laser light enters the optical cavity with a a well defined linear polarization. 
However, we have generally observed that this state changes inside the cavity.
This is correlated with the fact that the AR windows are not perfectly perpendicular to the beam. We confirmed this by deliberately changing the angle between windows and laser light and observing the corresponding change of the polarization.
The elimination of the AR windows inside the cavity should therefore solve this problem.

%%%%%%%%%%%%%%%%%%%%%%%%
%%%%%%%%%%%%%%%%%%%%%%%%%
 
%%%%%%%%%%%%%%%%%%%%%%%%%%%%%%%%%%%%%%%%%%%%%%%%%
%%%%%%%%%%%%%%%%%%%%%%%%%%%%%%%%%%%%%%%%%%%%%%%%%
%%%%%%%%%%%%%%%%%%%%%%%%%%%%%%%%%%%%%%%%%%%%%%%%%
\section{Data taking and analysis}\label{sec:analysis}

The data taking at the ALPS experiment proceeds in three steps. 
First, the regeneration tube is installed without the wall.
The small fraction of laser light transmitted through the central mirror (EM) attached to the end of the production tube is used to align the detector bench.
In the second step, the detector tube is removed, the wall is attached, the detector tube is reinstalled and evacuated and data taking takes place. 
After a certain period (typically one week) the detector tube is removed again, the wall detached and the open tube re-installed to check the alignment. 
This third step marks the end of one data taking period.

While data taking, one hour exposures are recorded by the CCD. 
Only frames for times where no technical problems (i.e. magnet quenches) occurred are kept for further analyses. 
These frames are checked for cosmics and other spurious tracks from radioactivity in a region of 25 by 25 pixels around the expected position of re-converted photons from WISPs. 
About 10\% of all one hour frames are rejected by this requirement.

Finally, re-converted photons are searched for by comparing the averaged sum of the ADU counts of all pixels in the signal region (see below) in data frames and dark frames. 
The latter ones are recorded with closed camera shutter and typically also with the magnet switched off and no or very low intensity laser light. Hence, any excess in the data frames could be a signature for WISP production.

It should be stressed that this simple approach holds true as long as no hint for an excess is derived from the data. If an excess is found, more detailed studies are to be performed demonstrating the WISP origin of the recorded photons.

\subsection{Signal search}\label{sec:databeamspot}

As sketched above, we used light-shining-through-the-mirror to align the detector optics and to define the search region for re-converted photons on the CCD. 
However, a systematic difference between the impact position of the light used for alignment and the light from photon-WISP-photon conversions arises from the non-perfect parallelism of front and back surface of the cavity end-mirror (EM). 
The traversing light is bent by the ``wedge'' shape of the mirror while the WISPs pass the mirror unaffected. 
We used two methods to determine the angle of the wedge:
\begin{itemize}
	\item The mirror is rotated and the diameter of the circular path of a distant beamspot is measured.
	\item The angular distance between the light passing through the mirror and the secondary reflection from the uncoated backside of the mirror is measured.
\end{itemize}
Both methods give an angle of 0.017° (0° for parallel front and back surfaces). With a distance of $\approx 8$~m between mirror and CCD and a lens in front of the CCD with a focal length of 60~mm one arrives at a difference of $\approx 8~\mu$m between the position of the beamspots from the light used for alignment and the location of the re-converted photons from WISPs. The CCD has a pixel size of $9~\mu$m, hence the search region is to be shifted by one pixel relative to the beamspot used for alignment (in our case upwards).
The analogous effects caused by the AR windows close to the EM (inside the cavity) and at the end of the detector vacuum tube were also tested. 
However, the resulting angles were found to be negligible.

\begin{figure}[t]	
\centering
\includegraphics[width=7.5cm]{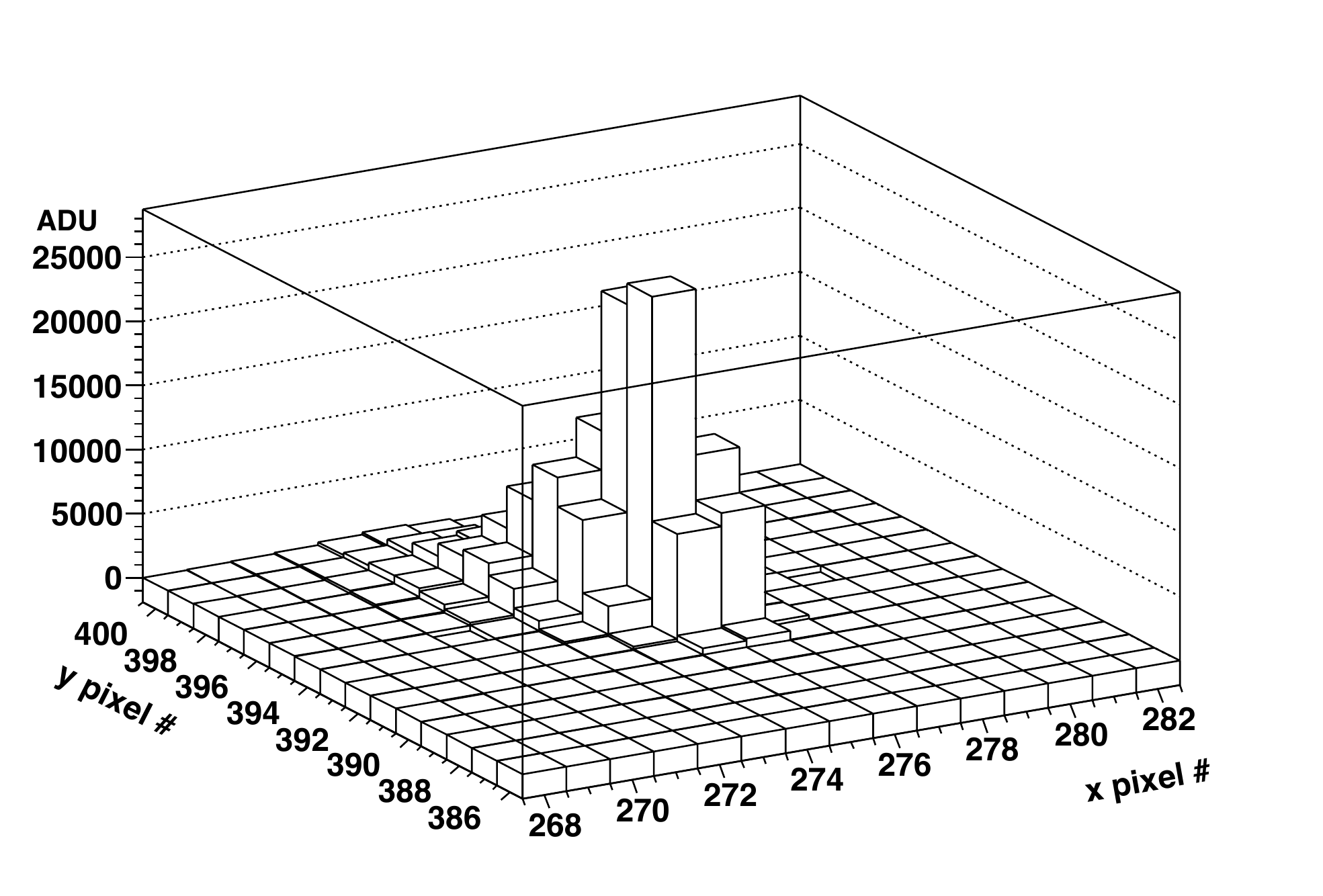}
\vspace{-0.3cm}	
\caption{The beamspot profile of the laser light leaking through the mirror in the centre of the ALPS magnet.}  
\label{fig:profile}
\vspace{-0.3cm}
\end{figure}

The beamspot profile of the re-converted photons is exactly the same as the beamspot used for alignment because the re-converted photons will have the same properties as the laser photons. The secondary reflection mentioned above will hit the CCD 5 pixels below the primary beamspot and is, as a conservative approach, not subtracted here because of its low intensity (less than 4\% of the primary spot). 
Therefore the signal profile used in this analysis is a little wider than the beamspot profile of re-converted photons from WISPs. 
Figure~\ref{fig:profile} shows the measured beamspot profile. 

%%%%%%%%%%%%%%%%%%%%%%%%%%%%
%%%%%%%%%%%%%%%%%%%%%%%%%%%%

Taking this beamspot profile and assuming a possible misalignment of $\pm$1 pixel in each direction (corresponding to the re-positioning accuracy mentioned above) at least 58\% of the signal is contained in a $5\times 5$ pixel large search region.

A hypothetical signal is searched for by comparing the mean value of the sum of the ADUs in the 25 pixels of the signal region in data frames and dark frames after correcting each frame for the pedestal fluctuations.

One principle difference in both frame sets is the presence of light from the reference beam in the data frames. The photon flux of the reference beam at the CCD is about 6.5~Hz ($2.4\times 10^{-18}$~W). By comparing dark frames and frames with the reference beam shining on the CCD we proved that the reference light does not affect the signal region. 
The distance between the signal region and the reference beam spot position amounts to 2~mm on the chip (230 pixels, see Fig.~\ref{fig:signal}).

After performing the pedestal correction, the dark frame data sets taken over the course of two years agree within statistical uncertainties. No long term drift or other changes are observed.

As a final check the dark frame data set (recorded in the course of 24 days) used for the analysis shown below is divided into a first and second half (104 frames each). Mean and width of the distribution of the sum of the pixels in the signal region agree very well within statistics, so that no hint on systematic differences is found.

\subsection{ALPS exemplary run}\label{sec:dataset}

The data set used for this analysis consists of 25 CCD frames with an exposure of one hour each. These frames were selected out of 31 frames taken during one week. Three frames, where a magnet quench occurred, are not taken into account for the data analysis. Another three frames are rejected, because the visual inspection of these frames shows hints for cosmic ray activity or radioactivity in the 25 by 25 pixel region around the position for the expected signal.

One hour long dark frames with the camera shutter closed were recorded to be compared with the data frames. After cutting against cosmics and radioactivity as described above 208 frames remain for the analysis.

The cavity internal power during data taking is shown in 
Fig.~\ref{fig:mr,cavity-internal power}. We achieved  
\begin{equation}
P_{\to}= 34 \pm \; \unit[4]{W} \ . 
\end{equation}
for the production of WISPs. 
The uncertainty originates from a conservative estimate of the systematic uncertainty of the calibration procedure, see Fig.~\ref{fig:calib}.

%%%%%%%%%%%%%%%%%%%%%%%%%%%%%%%%%%%%%%%%%%
\begin{figure}[t]	
\centering
\includegraphics[width=8.5cm]{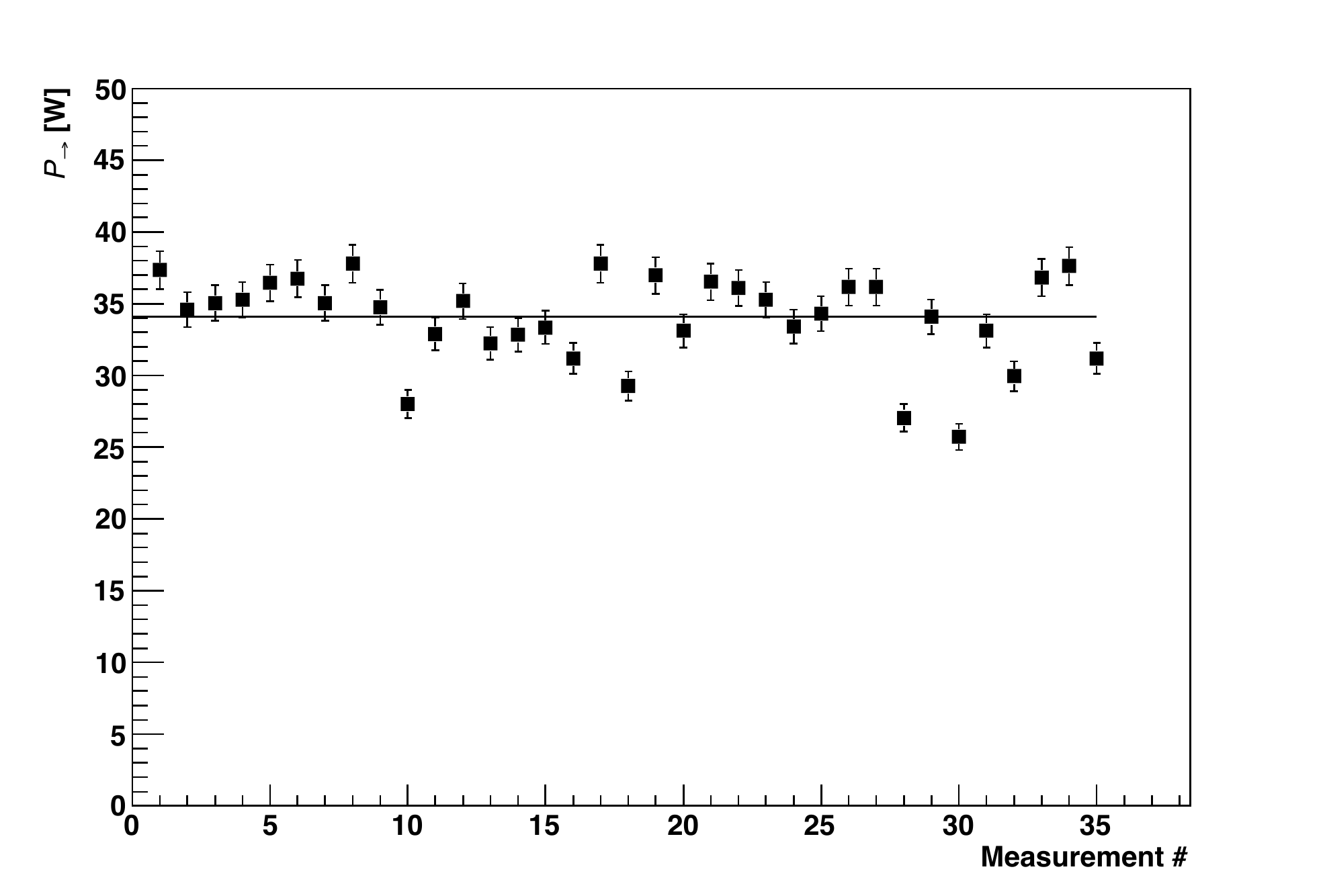}	
\vspace{-0.5cm}	
\caption{
Monitored power in the optical resonator during the collection of the data frames. 
The laser power was recorded each 30~s; only each hundredth point is shown for simplicity. 
The average power relevant for the measurements ($P_\to$) is determined to be $\unit[34\pm 4]{W}$. 
This corresponds to a power build-up of $43 \pm 5$ in excellent agreement with the characterization described in Sec.~\ref{sec:cav-characterization}.
\label{fig:mr,cavity-internal power}}
\vspace{-0.4cm}
\end{figure}
%%%%%%%%%%%%%%%%%%%%%%%%%%%%%%%%%%%%%%%%%%%

The linear polarization state of the eigenmode of the cavity during the data taking was measured (behind the mirror EM) to be 55 degrees with respect to the orientation of the magnetic field. We have thus 67$\pm$10\% of the laser photons with perpendicular polarization and 33$\pm$10\% with parallel polarization, see Table~\ref{tab:laspow}.

\begin{figure}[t]	
\centering
\includegraphics[width=9cm]{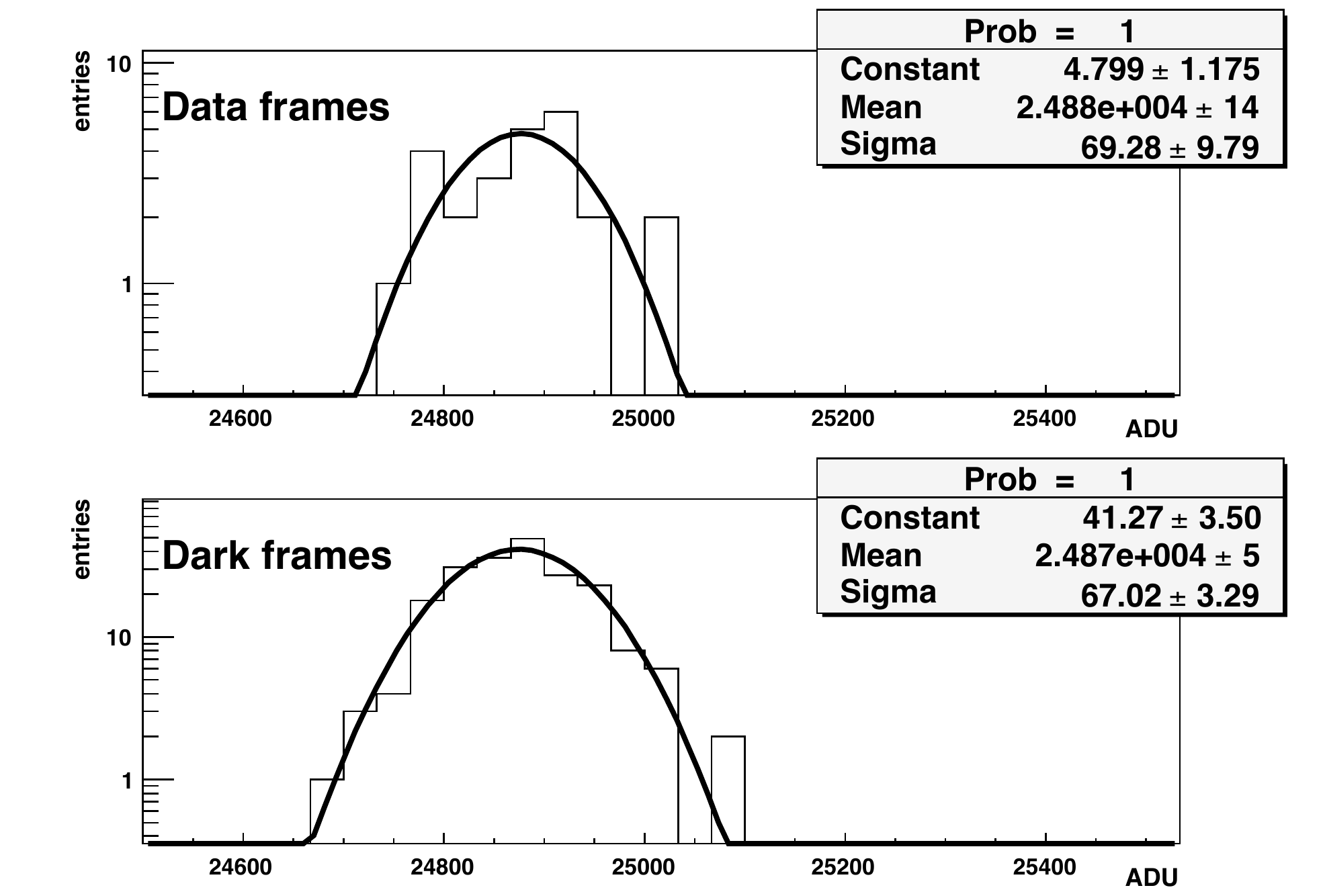}	
\caption{Distribution of the sums of the ADUs in the 5x5 pixels defining the search region for photons from re-converted WISPs. 
Each CCD frame (25 data frames, top, and 208 dark frames, bottom) gives one entry. The boxes show the results of Gaussians fitted to the distributions.}  
\label{fig:sumadu}
\vspace{-0.1cm}	
\end{figure}

From the alignment test before data taking the central position of the search region was determined to be at the (276,391) pixel coordinates of the CCD, while the alignment test after data taking gave the position (275,392). 
The difference by $\pm$1 pixels matches our experience for the re-positioning accuracy derived from numerous tests before. 
The distribution of the ADU sums in the search region is presented in Fig.~\ref{fig:sumadu} for both data and dark frame sets. 
Each frame is corrected for the pedestal fluctuation (see above) and gives one entry. 

%%%%%%%%%%%%%%%%%%%%%%%%%%%%
%%%%%%%%%%%%%%%%%%%%%%%%%%%%

%%%%%%%%%%%%%%%%%%%%%%%%%%%%
\begin{table}[tdp]
\centering
\begin{tabular}{cccc}
\hline
Polarization & Flux
\\[1pt] \hline 
Parallel & $(3.0\pm0.9)\times 10^{19}$Hz \\               
Perpendicular & $(6.1\pm0.9)\times 10^{19}$Hz  \\
Independent  & $(9.1\pm0.1)\times 10^{19}$Hz         \\
\hline
\end{tabular}
\caption{
Effective laser power relevant to the search for WISPs. 
Parallel, perpendicular and independent denote the orientation 
of the laser light polarization with respect to the magnetic field direction.}
\label{tab:laspow}
\end{table}
%%%%%%%%%%%%%%%%%%%%%%%%%%%%

\begin{figure}[t]	
\centering
\includegraphics[width=9cm]{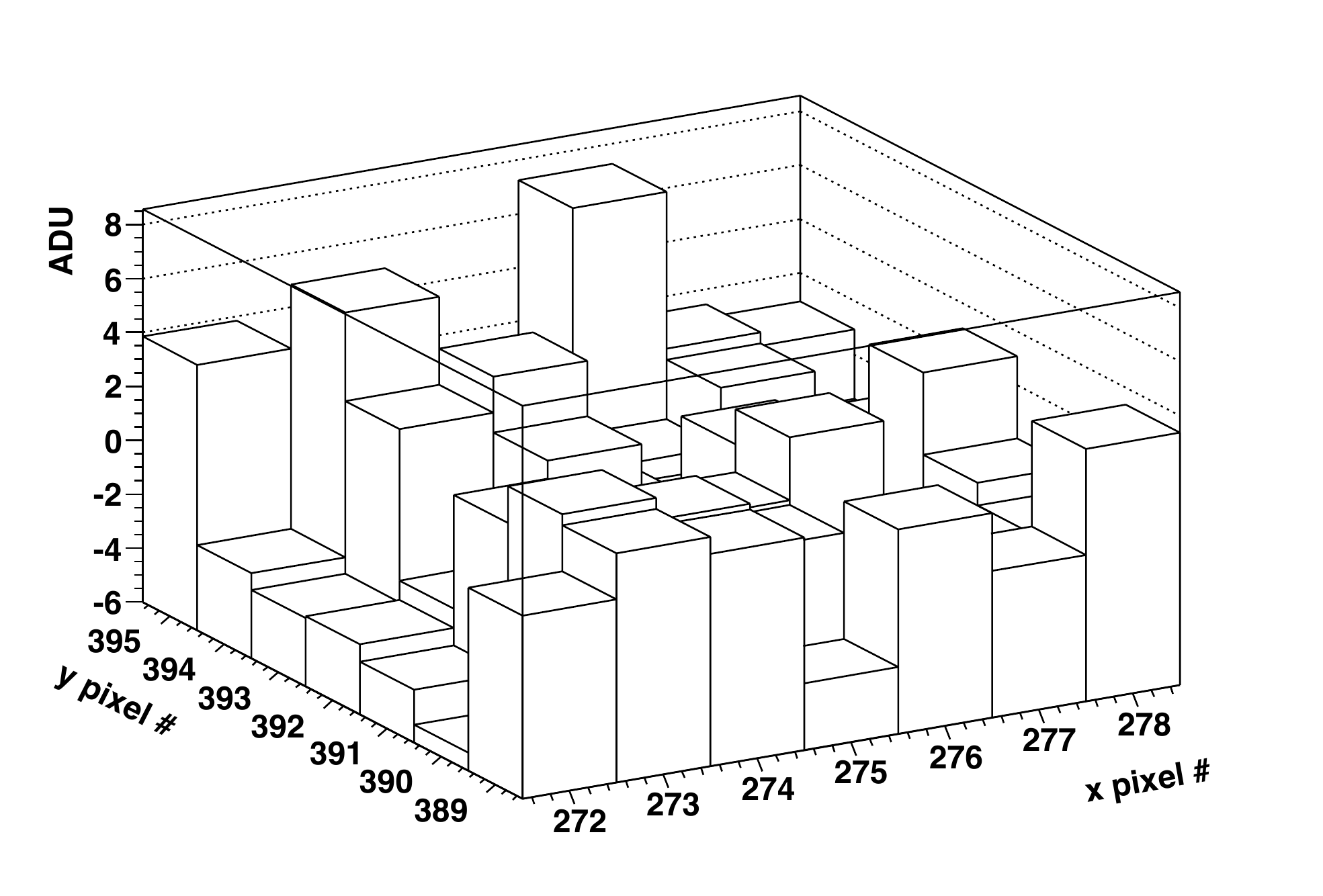}	
\vspace{-0.5cm}
\caption{
The difference between the signal (average of 25 data frames) and background (average of 208 dark frames)
ADUs shows no excess above noise expectations.
A signal of photons from re-converted WISPs should be centered around the position $(275\pm 1,392\pm 1)$.}  
\label{fig:adudiff}
\vspace{-0.1cm}
\end{figure}

It is obvious that there is no hint for any signal in the data frames: mean and width of both distributions agree very well within statistics. Also searching for a signal at other positions (see Fig.~\ref{fig:adudiff}) does not reveal any significant excess. Hence we do not observe any evidence for WISP production.

%%%%%%%%%%%%%%%%%%%%%%%%%%%%
%%%%%%%%%%%%%%%%%%%%%%%%%%%%

To derive upper limits for the WISP induced photon flux from our data we have to consider the quantum efficiency of the camera (0.6$\pm$0.1), the electron/ADU conversion factor of the camera (1.46$\pm$0.05) and the fraction of the signal contained in the search region (0.75$\pm$0.15).
This gives a flux limit of 28~mHz at 95\% C.L. (see Table~\ref{tab:fluxlimit}) using the method of \cite{Feld98}.
Taking into account the effective flux of incoming photons from Table~\ref{tab:laspow} one arrives at the conversion probabilities in Table~\ref{tab:convers}. 

%%%%%%%%%%%%%%%%%%%%%
\begin{table}[tdp]
\centering
\begin{tabular}{cccc}
\hline
ADU diff. & photon \# & flux & limit [95\% C.L.]\\[1pt] \hline
2.6$\pm$14.6 & 8$\pm$48 & (2$\pm$13)~mHz & 28~mHz 
\\
\hline
\end{tabular}
\caption{The difference of the average of the sums of ADU values of data and dark frames in the search region is shown in the first column. 
No significant excess is observed. 
The result is used to determine a 95\% C.L. flux limit for photons from re-converted WISPs.}
\label{tab:fluxlimit}
\end{table}
%%%%%%%%%%%%%%%%%%%%

\begin{figure}[ht]	
\centering
\includegraphics[width=9cm]{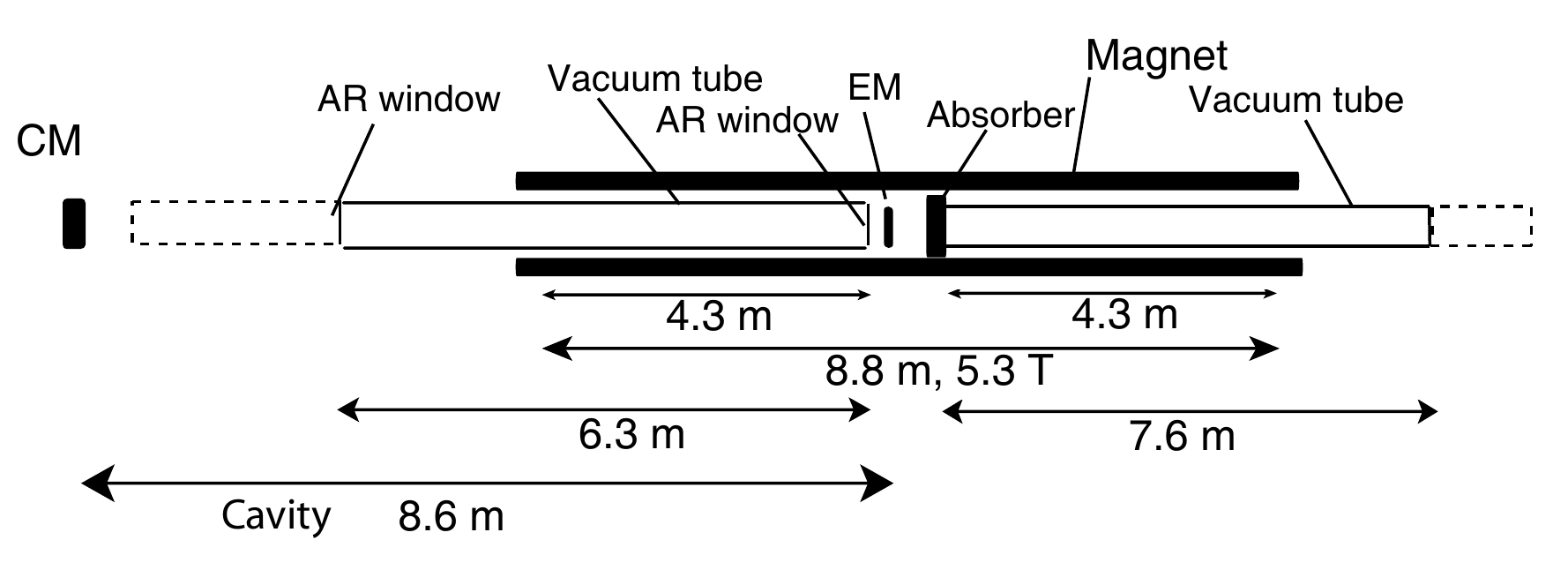}
\vspace{-0.5cm}	
\caption{Summary of the relevant experimental parameters for the ALPS WISP search. }  
\label{fig:summa}
\vspace{-0.3cm}
\end{figure}

For determining the limits on the photon-WISP mixing strength as a function of the WISP mass we use the experimental parameters summarized in Fig.~\ref{fig:summa}.
The uncertainties of the magnetic field strength, $\sigma_B$, (when relevant, i.e. for ALPs and MCPs) and of the tube lengths, $\sigma_L$, are taken into account by a Monte Carlo calculation. 
Here we conservatively use $\sigma_B =0.01$~T and $\sigma_L=0.1$~m. 
The results are shown in Fig.~\ref{fig:sens} and in Table~\ref{tab:sens}.

%%%%%%%%%%%%%%%%%%%%%%%%%%%%%%%%%%%%%%%%%%%%%%%%%
%%%%%%%%%%%%%%%%%%%%%%%%%%%%%%%%%%%%%%%%%%%%%%%%%
%%%%%%%%%%%%%%%%%%%%%%%%%%%%%%%%%%%%%%%%%%%%%%%%%
\section{Conclusions}\label{results}
%%%%%%%%%%%%%%%%%%%%%%%%%%%%%%%%%%%%%%%%%%%%%%%%%
%%%%%%%%%%%%%%%%%%%%%%%%%%%%%%%%%%%%%%%%%%%%%%%%%
%%%%%%%%%%%%%%%%%%%%%%%%%%%%%%%%%%%%%%%%%%%%%%%%%

The ALPS collaboration runs a light-shining-through-walls (LSW) experiment to search for photon oscillations
into ``weakly interacting sub-eV particles'' (WISPs) inside of a superconducting HERA dipole magnet at the site of DESY. 

In this paper we have described and characterized our apparatus and demonstrated the data analysis procedures.
Our main result is the first successful integration of a large-scale optical resonator into a complete LSW experiment. 
This resonator serves as an amplifier for the photon flux in the production region of the experiment and thus boosts the experiment's sensitivity. 

\begin{figure*}[ht]
\vspace{-0.6cm}
\centering
\includegraphics[width=14.4cm]{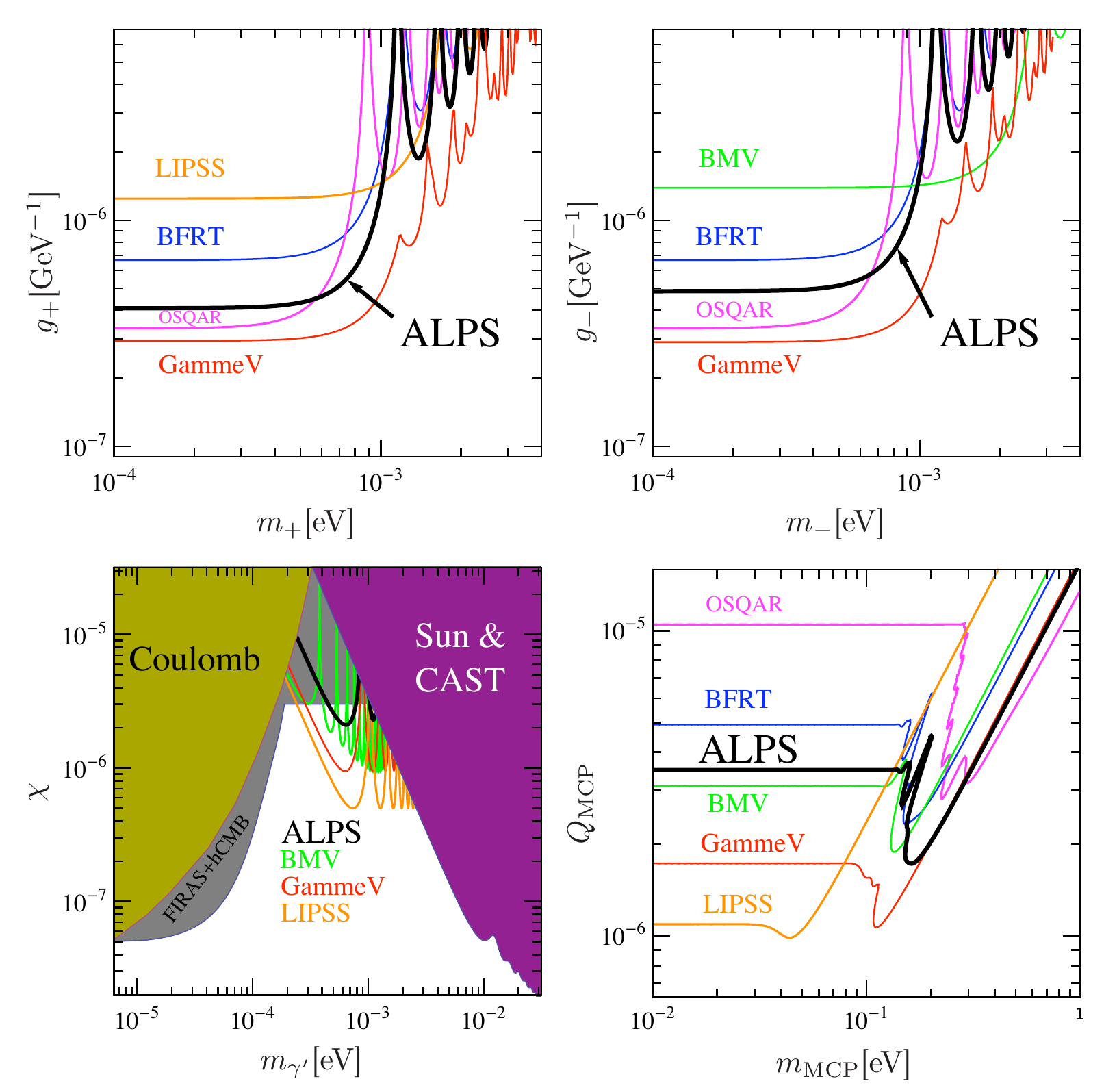}
\vspace{0cm}
\caption{Exclusion limits ($95\%$ C.L.) from the ALPS exemplary run (black). 
Starting from up-left corner and moving clockwise the limits state for scalar and pseudoscalar ALPs, 
mini-charged particles (in models with a massless hidden photon) and massive hidden photons. 
See Sec.~\ref{WISPs} for details and references. 
As a comparison we show exclusion limits of 
previous LSW experiments: BFRT~\cite{Cameron:1993mr}, 
BMV~\cite{Robilliard:2007bq,Fouche:2008jk}, GammeV~\cite{Chou:2007zzc,Ahlers:2007qf}, 
LIPSS~\cite{Afanasev:2008fv,Afanasev:2008jt} and OSQAR (preliminary)~\cite{osqarspsc:2007}. The results of a more recent OSQAR publication~\cite{Pugnat:2007nu} cannot be compared due to a different interpretation.
In all cases we present the bounds with $95\%$ C.L. 
In the hidden photon parameter space we have also plotted the  regions excluded from tests of the Coulomb law~\cite{Bartlett:1988yy,Williams:1971ms}, distortions of the cosmic microwave background and late formation of a hidden photon microwave background~\cite{Jaeckel:2008fi}, and emission of hidden photons from the Sun~\cite{Redondo:2008aa}.
Similar exclusion bounds for the case of ALPs and MCPs are review for instance in~\cite{Raffelt:1996wa,Redondo:2008en}.} 
\label{fig:sens}
\end{figure*}

During a 31 hours lasting exemplary run, the available laser light power to search for WISPs was increased by a factor of 43.
The upper limits on photon-WISP interactions derived from this brief run
show that ALPS is very competitive with other state-of-the-art LSW experiments
(see Fig.~\ref{fig:sens}).

Moreover, we have identified the main limitations of our current set-up, showing that significant improvements, especially of the performance of the optical resonator, are possible.  
This opens a clear path-way for near-term future steps in increasing the sensitivity of the ALPS experiment.

\begin{table}[h]
\centering
\begin{tabular}{cccc}
\hline
Polarization & Prob.$\times 10^{22}$ & 95\% C.L. & 99\% C.L. \\[1pt] \hline 
Parallel & $0.8\pm4.4 $ & $9.4$ & $12$ \\              
Perpendicular & $0.4\pm2.2$ & $4.5$ & $5.8$ \\
Independent & $0.3\pm1.4$ & $3.0$ & $3.9$ \\
\hline
\end{tabular}
\caption{
Probability of ``light-shining-through-walls'' tested with the ALPS exemplary run and corresponding upper limits.
Parallel, perpendicular and independent denote the orientation 
of the laser light polarization with respect to the magnetic field direction.}
\label{tab:convers}
\end{table}
\begin{table}[h]
\centering
\begin{tabular}{ccc}
\hline
WISP & 95\%C.L. Limit & Mass
\\[1pt] \hline 
ALP $0^-$ & $g<4.9\times 10^{-7}$ GeV$^{-1}$ & $m_-\lesssim$ 0.5~meV\\              
ALP $0^+$ & $g<4.1\times 10^{-7}$ GeV$^{-1}$ & $m_+\lesssim$ 0.5~meV \\ 
HP & $\chi < 2.1\times 10^{-6}$ & $\muu \approx$ 0.7~meV \\
MCP+HP & $Q_{\rm MCP} < 3.5\times 10^{-6}$ & $m_{\rm MCP}\lesssim$ 0.1~eV \\
\hline
\end{tabular}
\caption{Upper limits for the search for WISPs in the data presented in this paper.}
\label{tab:sens}
\end{table}

\newpage
%%%%%%%%%%%%%%%%%%%%%%%%%%%%
%%%%%%%%%%%%%%%%%%%%%%%%%%%%

\section*{Acknowledgments}
The ALPS collaboration thanks the MKS group at DESY for support, especially
H. Br\"uck, E. Gadwinkel, C. Hagedorn, H. Herzog, K. Pr\"uss and M. Stolper. 
Many thanks to U. Koetz for advice for selecting the camera and A. Zuber for the technical drawings.
We acknowledge financial support from the Helmholtz Association and from the Centre for Quantum Engineering and Space-Time Research (QUEST).

%\footnotesize
\bibliographystyle{utcaps} 
\bibliography{paper}

%\begin{thebibliography}{00}
%\bibitem{}
%\end{thebibliography}
\end{document}